\documentclass[usenatbib]{mnras}

\usepackage{amsmath,amssymb,gensymb}
\usepackage{multicol,tabularx}
\usepackage{enumitem}
\usepackage{graphicx,xcolor}

\usepackage[T1]{fontenc}
\usepackage{ae,aecompl}
\usepackage{txfonts}
\usepackage{color}
\usepackage{dsfont}
\usepackage{booktabs,multirow,subcaption,threeparttable}
\usepackage[capitalize]{cleveref}

\usepackage{etoolbox}
\makeatletter
\patchcmd\@combinedblfloats{\box\@outputbox}{\unvbox\@outputbox}{}{%
  \errmessage{\noexpand\@combinedblfloats could not be patched}%
}%
\makeatother

\newcommand{\twodots}{\mathinner {\ldotp \ldotp}} 
\newcommand{\subtext}[2]{\ensuremath{#1_{\text{#2}}}} 
\newcommand{\linel}[2]{#1\,$\lambda$#2} 
\newcommand{\kms}{\ifmmode {\rm km\,s}^{-1} \else km\,s$^{-1}$\fi} 
\newcommand{\diff}{\mathop{}\!\mathrm{d}} 

\title[Properties of BALQs to Illuminate Quasar Structure]{Using the Properties of Broad Absorption Line Quasars to Illuminate Quasar Structure}
\author[S.~Yong et al.]{\parbox[t]{\textwidth}{\vspace{-1cm}
Suk Yee Yong$^{1,}$\thanks{E-mail: \texttt{syong1@student.unimelb.edu.au}},
Anthea L.~King$^{1}$,
Rachel L.~Webster$^{1}$,
Nicholas F.~Bate$^{2}$,
Matthew J.~O'Dowd$^{3,4,5}$,
and Kathleen Labrie$^{6}$}
\vspace{1ex} \\
$^{1}$School of Physics, University of Melbourne, Parkville, VIC 3010, Australia\\
$^{2}$Institute of Astronomy, University of Cambridge, Madingley Road, Cambridge CB3 0HA, UK \\
$^{3}$Department of Physics and Astronomy, Lehman College of the CUNY, Bronx, NY 10468, USA\\
$^{4}$Department of Astrophysics, American Museum of Natural History, Central Park West and 79th Street, NY 10024-5192, USA\\
$^{5}$The Graduate Center of the City University of New York, 365 Fifth Avenue, New York, NY 10016, USA\\
$^{6}$Gemini Observatory, Hilo, HI 96720, USA
}%

\date{Accepted XXX. Received YYY; in original form ZZZ}
\pubyear{2018}

\begin{document}
\label{firstpage}
\pagerange{\pageref{firstpage}--\pageref{lastpage}}
\maketitle

\begin{abstract}
A key to understanding quasar unification paradigms is the emission properties of broad absorption line quasars (BALQs). The fact that only a small fraction of quasar spectra exhibit deep absorption troughs blueward of the broad permitted emission lines provides a crucial clue to the structure of quasar emitting regions. To learn whether it is possible to discriminate between the BALQ and non-BALQ populations given the observed spectral properties of a quasar, we employ two approaches: one based on statistical methods and the other supervised machine learning classification, applied to quasar samples from the Sloan Digital Sky Survey. The features explored include continuum and emission line properties, in particular the absolute magnitude, redshift, spectral index, line width, asymmetry, strength, and relative velocity offsets of high-ionisation \linel{\ion{C}{iv}}{1549} and low-ionisation \linel{\ion{Mg}{ii}}{2798} lines.

We consider a complete population of quasars, and assume that the statistical distributions of properties represent all angles where the quasar is viewed without obscuration. The distributions of the BALQ and non-BALQ sample properties show few significant differences. None of the observed continuum and emission line features are capable of differentiating between the two samples. Most published narrow disk-wind models are inconsistent with these observations, and an alternative disk-wind model is proposed. The key feature of the proposed model is a disk-wind filling a wide opening angle with multiple radial streams of dense clumps.
\end{abstract}


\maketitle

\section{Introduction} \label{sec:intro}

The optical-ultraviolet broad emission lines (BELs) are the most distinguishing feature of a quasar optical spectrum. These lines originate from the broad line region (BLR), which is situated close to the central ionising source (accretion disk). There have been various studies to understand the relationship between the geometry and dynamics of the BLR and the BEL profiles \citep[e.g.][]{Sulentic+:2000,Gaskell:2009,Grier:2013,Pancoast+:2014}. The absence of forbidden broad lines suggests that the BLR consists of dense regions that suppress the formation of these lines through collisional de-excitation of the gas. In addition, some quasars show broad absorption lines (BALs) blueward of the emission line. \citet{Chajet+Hall:2013,Chajet+Hall:2017,Yong+:2017,Braibant+:2017} have previously demonstrated that both the dynamics of the BLR emitting region and the angle of viewing of the quasar affect the shape and velocity offsets of the BELs. One of the main explanations for the BAL phenomenon and its rarity is a unification/orientation model that assumes all AGN possess BAL outflows but these outflows are narrow and BAL features are only observed when the observer's line-of-sight intersects the outflow. If the presence of a broad absorption feature results from a specific angle of viewing, the signature should be evident in the characteristics of the BELs. This is true no matter if the BLR is assumed to be co-spatial with BAL outflows, or not, or which BLR model (Keplerian disk, equatorial disk-wind, poloidal wind, etc.) is adopted. This study compares the measurable characteristics of BELs in quasars with and without BALs, to further understand the geometry of the BLR and BAL outflows.

Some clues to the nature of the BLR can be found by considering the unique properties of the BEL profiles. The shape of the BEL profiles and relative line strengths in a quasar spectrum vary from source to source and reflect the dynamics and geometry of the BLR. Typically, a BEL has full width at half maximum (FWHM) of $\sim 5000\,\kms$, but can be broader than 10\,000\,\kms\ \citep{Peterson:1997}. The emission line strengths vary between different line species in the same object or the same line in distinct objects, suggesting different ionisation conditions.

Often, emission line profiles are shifted relative to the systemic velocity of quasars. The forbidden [\ion{O}{iii}] narrow line is usually assumed to be within $\sim 50\,\kms$\ to the systemic velocity \citep{Hewett+Wild:2010}. Low-ionisation lines (LILs), such as \ion{Mg}{ii}, have average systemic velocities comparable to that of [\ion{O}{iii}] line, and hence are often used in cases where the [\ion{O}{iii}] line lies outside the observed wavelength range. A blueshift is also commonly observed between high-ionisation lines (HILs) and LILs. This velocity shift was first reported by \citet{Gaskell:1982}, and further confirmed by \citet{Wilkes:1986,Espey+:1989,Tytler+Fan:1992,McIntosh+:1999,VandenBerk+:2001,Shen+:2016}. HILs are also tend to be wider \citep[e.g.][]{Shuder:1982,Mathews+Wampler:1985}. This indicates that the HILs have different dynamics to the LILs and are likely to originate from a different parts of the BLR.

Results from reverberation mapping also show evidence that the BLR has a stratified ionisation structure \citep{Kollatschny:2003,Peterson+:2004}. The HILs and LILs are observed to arise at different relative distances from the inner accretion disk. Reverberation mapping studies find that the HILs such as \ion{He}{ii}, \ion{He}{i}, \ion{O}{vi}, \ion{N}{v} and \ion{C}{iv} are emitted from regions closer to the continuum source than the LILs such as \ion{Mg}{ii}, \ion{Ca}{ii}, \ion{O}{i}, H$\alpha$ and H$\beta$ \citep{Gaskell+Sparke:1986,Clavel+:1991,Peterson+Wandel:1999,Kollatschny:2003,Peterson+:2013,Bentz+:2016}. This reinforces the view that the location of the HIL regions are different from that of the LIL regions.

The emission lines display wide variation in shapes and degrees of asymmetry \citep{Corbin+Francis:1994,Corbin:1995,Corbin+Boroson:1996,Marziani+:1996,Corbin:1997}. Various studies relate line asymmetries, particularly for the \ion{C}{iv} and H$\beta$ lines, to radio types and orientation of quasars \citep{Marziani+:1996,Corbin:1997}. These studies find that in radio-loud sources the H$\beta$ line is redshifted and red asymmetric, while the \ion{C}{iv} line remains unshifted and symmetric. For radio-quiet sources, the \ion{C}{iv} line is blueshifted and blue asymmetric, while the H$\beta$ line is unshifted and symmetric.

Only a fraction of quasars are BALQs. A common measure of a BAL is the balnicity index \citep[BI;][]{Weymann+:1991}, which describes the amount of absorption blueward of the line:
\begin{align}
\text{BI}=\int^{-3000}_{-25\,000}\left[1-\frac{f(v)}{0.9}\right]C(v)\diff v,
\end{align}
where $f(v)$ is the normalised flux density as a function of velocity, $v$, relative to the centre of emission line. The value of $C$ is either 0 or 1. It equals to unity when the term in the square bracket is continuously positive for at least 2000\,\kms, and zero otherwise. Traditional BALs are defined to have BI $>0\,\kms$. Depending on the ionisation potential of the absorbed lines, BALQs are further separated into three subcategories: high-ionisation BAL (HiBAL) quasars, low-ionisation BAL (LoBAL) quasars, and iron low-ionisation BAL (FeLoBAL) quasars. HiBAL quasars show absorption from high-ionisation species, for example \ion{C}{iv}, \ion{N}{v}, and \ion{Si}{iv}, and are the most common type of BALQ. LoBAL quasars show absorption from high-ionisation species as well as absorption from low-ionisation species, such as \ion{Mg}{ii}, \ion{Al}{iii}, and \ion{Al}{ii}. FeLoBAL quasars are the rarest type of BALQ and are LoBALs with additional absorption lines from \ion{Fe}{ii} or \ion{Fe}{iii} complexes.

There are two primary explanations for the BAL phenomenon in quasars. One interpretation that we mentioned earlier is a unification model based on orientation. It is believed that all quasars have BAL outflows since the characteristics of the emission line and continuum of BALQs and non-BALQs are found to be similar \citep{Weymann+:1991}. The fact that only a small fraction $\sim 15$ per cent of quasars display BAL features \citep{Hewett+Foltz:2003,Reichard+:2003,Knigge+:2008,Gibson+:2009} might be due to orientation effects, such that a BAL is seen when the line-of-sight intersects the covering angle of the outflow. This leads to the notion of wind emanating from the accretion disk or the disk-wind model. The disk-wind model is often depicted as a biconical wind with a narrow opening angle of $\sim 10\degree\text{--}20\degree$ to account for the small fraction of BALQs, from which the BELs and BALs arise \citep{Murray+:1995,Elvis:2000,Elvis:2004}. However, for a narrow wind, we might expect different physical characteristics in the BELs between the BALQs and non-BALQs since they are observed from different directions \citep{Matthews+:2017,Yong+:2017}. For example, BALs viewed through a polar narrow wind will have narrower and more blueshifted line profiles than those of non-BALs observed from non-polar angles. In contrast, the line profile will be broad and less blueshifted in BALQs with equatorial narrow wind opening angle. We would expect different BEL properties for the two population when viewed from different orientations for any general flattened axisymmetric BLR models as shown by \citet{Collin+:2006,Goad+:2012} and \citet{Braibant+:2017}. An orientation explanation is not a new concept. Orientation has also been commonly used in the literature to differentiate between the types of active galactic nucleus (AGN), either type 1 or type 2, and based on radio morphology, either radio-loud or radio-quiet \citep{Antonucci:1993,Urry+Padovani:1995}.

An alternative interpretation for the BAL phenomena is that the BALQ represents a stage in the evolution of quasars. In this scenario, BALQs are young quasars residing in a gas- and dust-rich environments enveloped by a high covering fraction cocoon \citep[e.g.][]{Hamann+Ferland:1993,Voit+:1993,Becker+:2000}. After some time they blow off their dusty shroud and become the more common non-BALQs. The spectrum of a BALQ is found to be redder compared to a non-BALQ spectrum with LoBAL quasars redder than HiBAL quasars \citep{Weymann+:1991,Sprayberry+Foltz:1992,Brotherton+:2001,Reichard+:2003,Trump+:2006,Gibson+:2009}, suggesting a quasar transition from FeLoBALs, LoBALs, HiBALs, and finally to non-BALs. If the BAL phenomenon is an evolutionary stage of a quasar, then we might expect no difference between BEL characteristics of the two populations, but a difference in spectral slope and maybe differences in accretion rates and/or black hole masses. It is also possible that the BAL phenomenon is a combination of both explanations \citep{Gallagher+:2007,Allen+:2011,DiPompeo+:2013}.

In a recent paper by \citet{Matthews+:2017}, the distribution of the BEL equivalent width (EW) values for both BALQ and non-BALQ are compared to test the geometric unification model. They find similarities between the EWs of both populations, which contradicts the idea that a BAL originates from an equatorial outflowing wind coming off a geometrically thin but optically thick accretion disk. These authors conclude that either (i) the continuum emission is inconsistent with a geometrically thin accretion disk, (ii) the viewing angles for BALQ and non-BALQ are the same (i.e., at low inclination angles close to face-on), or (iii) geometric unification is unable to justify the BAL fraction in quasar samples.

In \citet{Yong+:2016,Yong+:2017}, we developed a simple kinematical thin disk-wind model, enabling a qualitative understanding of both the emission line widths and the offsets of line centroids. The results of those analyses demonstrate that the properties of the BELs are highly dependent on the viewing angle, wind opening angle, and wind region. \citet{Yong+:2017} find that the shape of the emission line profile is narrow and asymmetric when viewed face-on. In contrast, the emission line is broad and symmetric for an edge-on geometry. The relative blueshift is larger as the line-of-sight is aligned with the outflowing wind. Additionally, a polar narrow wind model exhibits higher blueshift compared to that of equatorial wind model. The aim of these investigations was to relate the measured attributes of the BELs to the observed in the context of a disk-wind model for the BLR.

Investigating the differences in emission properties between the BALQ and non-BALQ populations can therefore shed light on the geometry and origin of the observed BALs. In this paper, we attempt to group the BAL and non-BAL quasars into their respective classes using statistical tests and supervised machine learning for classification. The investigated features include continuum properties, particularly the absolute magnitude, redshift, and spectral index, and additionally the characteristics of high- and low-ionisation emission lines, specifically the FWHM, asymmetry, EW, and velocity offsets of \linel{\ion{C}{iv}}{1549} and \linel{\ion{Mg}{ii}}{2798} emission lines. We explicitly consider whether the differences between the BALQs and the non-BALQs can be explained by either of the postulated paradigms, and if so, what constraints are required on the model. Throughout this analysis, we assume that the BELs and BALs arise in a consistent model of the disk-wind and we refer to this region as the BLR.

The outline of the paper is as follows. In \cref{sec:dataset}, we describe the selection criteria for the data sample of quasars and separate them into BAL and non-BAL populations. The statistical tests and machine learning algorithms employed are outlined in \cref{sec:method}. The results from the two methods are presented in \cref{sec:results}, followed by discussion in \cref{sec:discussion} where we examine the use of the observable BAL signatures in the context of evolutionary and orientation in a narrow disk-wind paradigms. This analysis motivates us to present a revised BLR model in \cref{sec:proposedblr}. Finally, a summary is provided in \cref{sec:summary}.

\section{Dataset} \label{sec:dataset}

The data used in this paper are obtained from the Baryon Oscillation Spectroscopic Survey \citep[BOSS;][]{Dawson+:2013} of the Sloan Digital Sky Survey III \citep[SDSS-III;][]{Eisenstein+:2011} Data Release 12 Quasar \citep[DR12Q;][]{Paris+:2017} catalogue. The SDSS DR12Q catalogue is retrieved through the VizieR catalogue access tool\footnote{\url{http://vizier.u-strasbg.fr/viz-bin/VizieR}} \citep{Ochsenbein+:2000}, which is publicly available online. The details of the estimation of the global redshift, emission line redshift, and subsequently the widths and the rest frame EW for each line, are described in Sect.~4 of \citet{Paris+:2012}. We will briefly summarise them here.

\citet{Paris+:2017} first estimated the overall redshift via visual inspection, \subtext{z}{vi}. The spectra are then fitted using a linear combination of four principal components constructed from a high quality subset of SDSS DR7Q \citep{Schneider+:2010} spectra that do not exhibit BAL features. The redshifts are adjusted until the best fit is found, which yields the principal component analysis (PCA) redshift, \subtext{z}{PCA}. The individual emission lines are then fitted by a set of five principal components. The redshift of an emission line is estimated from the peak amplitude of the fitted line. The velocity offset between the lines is computed as the difference in the individual emission line redshifts. The symmetry of the line is defined as the ratio of the blue to red half width at half maximum (HWHM), which are evaluated bluewards and redwards of the emission line peak, respectively. The FWHM is then the sum of both HWHM values.

We adopt the convention that a blueshifted line has a negative velocity and will peak towards the bluer end of the wavelength. Conversely, the peak of a redshifted line has a positive value and tends towards longer wavelengths. A negative or blueward asymmetric line will have more flux on the blue wing, while a positive or redward asymmetric line will have a broader red wing.

\citet{Paris+:2017} implemented the BALQ definition from \citet{Weymann+:1991} to identify BAL feature in \ion{C}{iv} line. For a line to be considered as a BAL, the BI needs to be larger than 0\,\kms, have an absorption trough width of $\geq 2000\,\kms$ at 10 per cent depth below the continuum, and be blueshifted $\geq 3000\,\kms$ from the emission line. The \ion{C}{iv} BAL troughs are automatically detected for quasars with $\subtext{z}{vi} \geq 1.57$ to ensure that the region from \ion{S}{iv} to \ion{C}{iv} is included. The continuum is estimated using a linear combination of four principal components fit to the spectra iteratively to mask any absorption in the spectrum \citep[see examples in Fig.~15 of][]{Paris+:2012}. Based on the constructed continuum, the BI in the blue side of \ion{C}{iv} emission line is calculated.

The flux density of the continuum can be represented by a power law, given by $\subtext{f}{cont} \propto \nu^{\alpha_{\nu}}$, where $\nu$ is the frequency. The spectral index, $\alpha_{\nu}$, is retrieved by applying this approximation to fit over regions with rest wavelength of 1450--1500, 1700--1850, 1950--2750\,\AA. These regions are selected as they are free from emission lines. Based on the SDSS primary photometry, the $i$-band absolute magnitude at redshift $z=2$, $M_{i}(z=2)$, is computed assuming $\alpha_{\nu}=-0.5$ and $K$-correction from Table~4 in \citet{Richards+:2006}.

We select a subsample for this analysis such that both \ion{C}{iv} and \ion{Mg}{ii} emission lines, and \ion{C}{iv} BAL features are present in the spectrum. For this purpose, the redshift is bounded for \subtext{z}{vi} and $\subtext{z}{PCA} \geq 1.57$. The sample was also chosen to ensure that all measurements had high signal-to-noise, so that results are not biased by poor statistics. Thus the sample is further constrained to ensure that the spectra have prominent emission lines: (i) median signal-to-noise ratio (S/N) per pixel over the whole spectrum $\geq 15$, (ii) \ion{C}{iv} and \ion{Mg}{ii} FWHMs $>0\,\kms$, and (iii) \ion{C}{iv} amplitude $\geq 10$ and \ion{Mg}{ii} amplitude $\geq 5$ median rms pixel noise. Although the \ion{C}{iii}] line is also included in this wavelength range, the uncertain contributions of \ion{Al}{iii} and \ion{Si}{iii}] emissions to the wings of the line suggest that an analysis including this line may not be as robust unless the \ion{Al}{iii} and \ion{Si}{iii}] emissions are deconvolved.

We remove the data with missing or undefined $\alpha_{\nu}$ values from the raw data. Three outliers are also taken out from the sample. One has $M_{i}(z=2) \sim -45$, much brighter than the rest of the samples which have $M_{i}(z=2)<-35$. Upon visual inspection, we found that the other two with either EW(\ion{Mg}{ii}) $<0$ or blue/red HWHM(\ion{Mg}{ii}) $>10\,000$, have inaccurate measurements. The final sample consists of 2773 spectra and among those 313 show a BAL feature for the \ion{C}{iv} line. This corresponds to a BAL quasar fraction of $\sim 11.29$ per cent. Hereafter, a BALQ refers to a quasar with the presence of a \ion{C}{iv} line HiBAL feature only, unless mentioned otherwise. LoBALs might be present but are not identified in the dataset.

\section{Methodology} \label{sec:method}

Using the dataset presented in the previous section, we conduct several tests to find out whether the BALQ and non-BALQ populations belong to the same parent population using continuum and BEL properties. The two approaches employed are based on statistics and machine learning techniques, which will be described in \cref{ssec:statstest} and \cref{ssec:ml} respectively. A total of 12 features are investigated in this work, as listed in \cref{tab:features}.

\begin{table*}
\centering
\caption{List of investigated features using continuum and broad emission line properties.}
\label{tab:features}
  \setlength{\tabcolsep}{3pt}
  \begin{threeparttable}
  \begin{tabular}{@{\extracolsep{0pt}}l l l c@{}}
    \toprule
    Property & Feature & Description & Notation \\
    \midrule
    Continuum & \texttt{imag} & Absolute magnitude in $i$-band at $z=2$ & $M_{i}(z=2)$ \\
     & \texttt{z.pca} & PCA redshift & \subtext{z}{PCA} \\
     & \texttt{alphanu} & Spectral index & $\alpha_{\nu}$ \\
    \cline{1-4}
    BEL & \texttt{fw(civ)} & FWHM of \ion{C}{iv} & FWHM(\ion{C}{iv}) \\
     & \texttt{civ\_ratioskew}\tnote{*} & Asymmetry of \ion{C}{iv} & Blue/red HWHM(\ion{C}{iv}) \\
     & \texttt{w(civ)} & EW of \ion{C}{iv} & EW(\ion{C}{iv}) \\
     & \texttt{fw(mgii)} & FWHM of \ion{Mg}{ii} & FWHM(\ion{Mg}{ii}) \\
     & \texttt{mgii\_ratioskew}\tnote{*} & Asymmetry of \ion{Mg}{ii} & Blue/red HWHM(\ion{Mg}{ii}) \\
     & \texttt{w(mgii)} & EW of \ion{Mg}{ii} & EW(\ion{Mg}{ii}) \\
     & \texttt{civmgii\_diffv}\tnote{*} & Velocity offsets of \ion{C}{iv} and \ion{Mg}{ii} & $\Delta v$(\ion{C}{iv}-\ion{Mg}{ii}) \\
     & \texttt{civmgii\_ratiofwhm}\tnote{*} & FWHM ratio of \ion{C}{iv} and \ion{Mg}{ii} & FWHM(\ion{C}{iv}/\ion{Mg}{ii}) \\
     & \texttt{civmgii\_ratioew}\tnote{*} & EW ratio of \ion{C}{iv} and \ion{Mg}{ii} & EW(\ion{C}{iv}/\ion{Mg}{ii}) \\
    \bottomrule
  \end{tabular}
  \begin{tablenotes}[flushleft]\footnotesize\setlength{\labelsep}{0pt}
    \item {\bfseries Note:} Features are as those defined in VizieR access tool, except stated otherwise.
    \item[*] User defined features from SDSS measurements.
  \end{tablenotes}
  \end{threeparttable}
\end{table*}

\subsection{Statistical Tests} \label{ssec:statstest}

To determine whether two samples are likely to be drawn from the same population, a two-sample (2s) statistical test based on the empirical distribution function (EDF) can be conducted. The two well-known EDF tests are Kolmogorov--Smirnov \citep[K--S;][]{Kolmogorov:1933,Kolmogorov:1941,Smirnov:1939} and Anderson--Darling \citep[A--D;][]{Anderson+Darling:1954,Darling:1957,Pettitt:1976} tests. Both are non-parametric statistical methods that are free from any assumption about the probability distribution of the data and test the null hypothesis, $H_{0}$, that the samples belong to the same distribution.

For the K--S test, a two-tailed $p$-value is evaluated. We describe the statistics as highly significant at $<0.1$ per cent, significant at $<5$ per cent, and not significant at $>5$ per cent. The $p$-value can be interpreted as follows. If the $p$-value is $\leq 5$ per cent, there is sufficient evidence to reject the null hypothesis, suggesting that the two samples are not chosen from the same distribution. On the other hand, if the $p$-value is $>5$ per cent, there is not enough evidence to reject the null hypothesis.

Compared to the K--S test, the A--D test is more sensitive to the changes at the tails of the distributions. The calculated significance level, $\alpha$, is the probability of the null hypothesis being rejected when it is true. Since the interpretations for $p$-value and $\alpha$ are similar, we adopt the same statistical significance cut-off for both.

The 2s statistical tests are computed using an open source scientific tools for Python, \texttt{scipy}\footnote{\url{https://scipy.org/}} \citep{Jones+:2001}. The tests are calculated for one-dimensional (1d) case between BALQ and non-BALQ datasets.

\subsection{Machine Learning} \label{ssec:ml}

The goal of using machine learning (ML) is to find the parameter space that plays the biggest role in separating the BALQ and non-BALQ populations without human intervention, providing us some clues to the dynamics and origins of the BLR. For the purposes of this study, we use supervised ML for binary classification problems and only review these methods. The choice of ML algorithms is restricted to those that return feature importance or weighting such that the influence of each features can be quantified. We start with four basic supervised ML classification algorithms for interpretability, namely: decision tree, random forest, logistic regression, and support vector machine.

\subsubsection{Algorithms}

\paragraph{Decision Tree} ~\\
Decision tree \citep{Breiman+:1984} is a predictive model that uses a series of observations and sample splitting according to those observations to make conclusions about the possible classification of an input. The training process starts at the tree root node, i.e., the origin of all branches, and the samples are split into branches of child nodes. At a node, all features are considered and the split is chosen that maximises the purity of the children sample given the parent sample. The purity of the sample can be thought of the least amount of mixing between the different classifications. In the case, where the parent node is a mix of objects with a classification of either A or B, the perfect split (aka the split that gives the maximum purity) would perfectly split the sample with classification A into one child node and objects with classification B into the other child node. This purity can rarely be achieved after one split. Therefore, the splitting procedure is iterated at each child node until the samples are perfectly separated or upon reaching the specified minimum number of samples at the node. The final node is called the terminal or leaf node, and the predicted outcome is given at this point.

Technically, the objective function of the decision tree algorithm is to maximise the impurity decrease at each split. The impurity decrease, $\Delta I(s,t)$, evaluates the quality of a split $s$ given node $t$. In the binary case, the parent node at $t$ is separated into two child nodes, the left (\subtext{t}{L}) and the right (\subtext{t}{R}):
\begin{align}
\Delta I(s,t)=I(t)-\frac{N_{\subtext{t}{L}}}{N_{t}}I(\subtext{t}{L})-\frac{N_{\subtext{t}{R}}}{N_{t}}I(\subtext{t}{R}),
\label{eqn:impuritydecrease}
\end{align}
where $N_{t}$, $N_{\subtext{t}{L}}$, and $N_{\subtext{t}{R}}$ are the number of samples in the parent, left, and right nodes, respectively. The quantity $I(t)$ is the purity in the sample and $\Delta I(s,t)$ describes the changes in purity from the parent node to the left and right child nodes.

The impurity measure, $I(t)$, is evaluated using Gini index \citep{Gini:1921}, which minimises the probability of misclassification. For binary classification, it is defined as
\begin{align}
\subtext{I}{G}(t)=2p(i|t)[1-p(i|t)].
\end{align}
The parameter $p(i|t)$ is the fraction of samples that are in class $c$ at node $t$. The impurity decrease based on Gini index, $\Delta \subtext{I}{G}(s,t)$, is also known as Gini impurity.

\paragraph{Random Forest} ~\\
Random forest \citep{Breiman:2001} is an ensemble of decision trees. A collection of trees is built from random samples generated by bootstrap sampling without replacement. The final classification is obtained via averaging all the votes from each trees. By combining multiple independently trained decision trees, the variance captured by individual trees in the forest is reduced.

\paragraph{Logistic Regression} ~\\
Logistic regression \citep{Cox:1958}, also called logit regression or maximum-entropy classification, belongs to a generalised linear model. It attempts to search for the optimal hyperplane in the input quasar features (see \cref{tab:features}) that separates the sample into BALQs and non-BALQs by maximising the log likelihood function. The algorithm is used to predict the probability of being one of two possible classifications based on the input features. The probability model is characterised by a logistic or sigmoid function
\begin{align}
P(y=\pm 1|\vec{x},\vec{w})=\frac{1}{1+\exp[-y(w_{0}+\vec{w}^{T}\vec{x})]},
\end{align}
where $\vec{x}$ are the predictor variables with the corresponding weight vector $\vec{w}$, and $w_{0}$ is the intercept or bias. The value is compared with a threshold of 0.5 to determine either the object is a BALQ or a non-BALQ. If the likelihood is greater than the threshold, the object will be predicted as a BALQ, and vice versa if it is less. The weight function is iteratively calculated using a training sample and is chosen to minimise the deviations between the predicted classification given a trial weight function and the known classification.

\paragraph{Support Vector Machine} ~\\
Support vector machine \citep[SVM;][]{Boser+:1992,Cortes+Vapnik:1995} is a discriminative classifier with the aim of obtaining a maximum margin hyperplane that divides the classes. Margin is defined as the distance between the closest points, the support vectors, and the hyperplane. The hyperplane is a linear combination of input vector, $\vec{x}$, and is written as
\begin{align}
f(\vec{x})=w_{0}+\vec{w}\cdot\phi(\vec{x}),
\end{align}
where $\vec{w}$ is the weight vector, $w_{0}$ is the intercept, and $\phi$ is a mapping function to a feature space. For the weight to have meaning in terms of feature importance, we restrict the kernel type of the SVM algorithm to a linear function. The algorithm searches for decision boundaries with the largest margin and solves for the hyperplane parameters. Once the optimal hyperplane is found, discrete class labels are assigned to make predictions on the new points.

\subsubsection{Building Machine Learning Classifiers}

We attempt to construct a supervised binary classification algorithm that can differentiate between the two populations, using a set of continuum and emission line properties. We examine if there is any one of them that outperforms in dividing the samples into their respective classes. The analyses use \texttt{scikit-learn}\footnote{\url{http://scikit-learn.org/}} \citep{Pedregosa+:2011}, an open source ML package in Python. Each of the algorithms utilised in this analysis requires tuning via hyperparameters. For example, the number of decision tree splits for a given decision tree, the number of decision trees used in the the random forest algorithm, or the regularisation of the logistic regression algorithm that controls the complexity vs.\ fit balance when finding the optimal weight vector. An outline of the steps performed is as following:
\begin{enumerate}
  \item Data preprocessing (Section~\ref{para:dataprocess})
  \begin{itemize}
    \item Load raw data and remove those with missing values and bad measurements. Separate the data into binary classes of BALQ and non-BALQ populations based on BI values, where BALs are those with BI $>0\,\kms$ in \ion{C}{iv} line.
    \item Split data into 80 per cent training and 20 per cent test samples.
    \item Apply feature scaling to dataset.
  \end{itemize}
  
  \item Tuning the models (Section~\ref{para:tunemodel})
  \begin{itemize}
    \item Conduct coarse grid searches on algorithm hyperparameters with stratified 10-folds cross-validation (CV) using training set and evaluate the performance using scoring metrics. Stratified 10-folds CV will be described in Section~\ref{para:tunemodel}.
    \item Conduct fine grid search and randomised search on algorithm hyperparameters with stratified 10-folds CV.
  \end{itemize}
  
  \item Model evaluation (Section~\ref{para:modelevaluation})
  \begin{itemize}
    \item Obtain the best estimator with the highest validation score.
    \item Compare the performance of the different models using various scoring metrics.
    \item Predict the classes of the test set. Extract the feature importances and compare their significance in other models.
  \end{itemize}
\end{enumerate}

\paragraph{Data Preprocessing} \label{para:dataprocess} ~\\
After the samples have been refined (see \cref{sec:dataset}), the dataset is partitioned into 80 per cent training and 20 per cent test samples. The training set consists of quasars from SDSS DR12Q catalogue, with known classes. It is used to learn a model and make a prediction on the new data. The performance of the model is then assessed using the test set.

Logistic regression and SVM algorithms require the data to be scaled, while tree-based estimators are scale-invariant. Since there are possibly some outliers, a robust scaler is used to centre the median to zero and scales the input parameters by the interquartile range, which is the difference between the upper (75th) and lower (25th) quartiles.

An imbalanced dataset may render model predictions inaccurate towards the more common class. In the dataset, non-BALQs are more abundant and the estimator will attempt to maximise the prediction for this class and overlook the BALQ sample. To account for the imbalanced population, the weighting for each class is set to be the same. As a consequence, the cost of misclassifying the minority class is increased and the classifier will assign equal emphasis on predicting both classes correctly. A scoring metric that is sensitive to imbalanced classes is also employed to alleviate this issue as described next.

\paragraph{Scoring Metrics} ~\\
One way to illustrate the number of samples that are correctly predicted in each class is with a contingency table or confusion matrix. Here, the binary classes can be denoted as non-BALQ (negative) and BALQ (positive). The resulting confusion matrix for binary classification is displayed in \cref{tab:cmatrix}. A negative class sample that is correctly labelled is referred to as true negative (TN), while false positive (FP) if it is misclassified as positive. A sample in positive class that is predicted correctly is called true positive (TP), while false negative (FN) when it is wrongly identified as negative.

A scoring metric is assigned to quantify the efficacy of the model predictions. Model evaluations that use accuracy are unreliable in the presence of imbalanced dataset. Instead, a F1 score is used as the scoring parameter as it is slightly more sensitive to class imbalanced data. The F1 score represents the balance between precision and recall, given by
\begin{align*}
\text{F1}=2\left(\frac{1}{\text{precision}} + \frac{1}{\text{recall}}\right),
\end{align*}
where precision=TP/(TP+FP) is the probability of predicting positive instances and recall=TP/(TP+FN) or sensitivity is the probability of detecting positive instances. The score ranges between 0 and 1, with 1 indicating perfect precision and recall.

\begin{table}
\centering
\caption{Confusion matrix for binary classification.}
\label{tab:cmatrix}
  \begin{tabular}{@{\extracolsep{4pt}}c c c c@{}}
    \toprule
     & & \multicolumn{2}{c}{Actual} \\
     \cline{3-4}
     & Class & non-BALQ & BALQ \\
    \midrule
    \multirow{4}{*}{\rotatebox[origin=c]{90}{Predicted}} & non-BALQ & True & False \\
     & & negative (TN) & negative (FN) \\
     & BALQ & False & True \\
     & & positive (FP) & positive (TP) \\
    \bottomrule
  \end{tabular}
\end{table}

Subsets of the training dataset can be further split using cross-validation (CV) technique to tune the variables. One of CV strategies is a stratified k-fold CV, which is a sampling method that retains the initial fraction of each class. This method has been shown to reduce the bias and variance estimation in imbalanced data \citep{Kohavi:1995}. The strategy is as follows. The training set is randomly partitioned into $k$ folds without replacement, i.e., each sample will be selected exactly once. Then, $k-1$ subsets are used to train the data and one subset for testing the performance. The one remaining fold is called the holdout or validation set. The preceding step is repeated $k$ times. Lastly, all the model performances are averaged.

\paragraph{Tuning the Models} \label{para:tunemodel} ~\\
Before initialising a ML algorithm, there are a few variables that need to be specified. These variables are termed the hyperparameters and require tuning by exploring the parameter space. To calibrate the hyperparameters for a particular algorithm, parameter space searches are performed with stratified 10-folds CV. In general, there are two strategies to conduct a parameter search: grid and randomised search. Grid search CV explores a set of predefined parameter ranges. Randomised search CV probes parameters sampled from a prescribed distribution for a specified number of candidates. We perform both strategies to check the consistencies of the searches.

As outlined earlier, a coarse grid search with stratified 10-folds CV is first executed to find an appropriate range of hyperparameter values. The performance of the models is evaluated using a F1 score. From this investigation, a set of parameter ranges is assembled for a fine grid search. Using the same range of parameter space, a randomised search is also sampled for 1000 iterations.

The three hyperparameters that are tuned in decision tree classifier are
\begin{itemize}
  \item \texttt{max\_depth}: Maximum depth of tree.
  \item \texttt{max\_features}: Maximum features considered in search for the best split.
  \item \texttt{min\_samples\_leaf}: Minimum number of samples in a leaf node.
\end{itemize}
Although there are other hyperparamters that can be calibrated, they are of secondary importance, and hence are omitted to save computational time.

Since a random forest classifier is a collection of decision trees, the hyperparameters are the same as those used in the decision tree analysis. Following the procedure in the previous section, \texttt{max\_depth}, \texttt{max\_features}, and \texttt{min\_samples\_leaf} are tuned by grid search. An additional hyperparameter, which is the number of trees, \texttt{n\_estimators}, can also be adjusted. A higher number of trees reduces the variance in the model but with the cost of increasing computational time. Due to this, the number of trees is fixed to be 100.

The logistic regression algorithm is implemented using a library for large linear classification (LIBLINEAR)\footnote{\url{https://www.csie.ntu.edu.tw/~cjlin/liblinear/}}. The grid searches on the hyperparameter $C$, which controls the regularisation, are done in a sequence of 100 evenly spaced samples in logarithmic scale. The internal computation for SVM in \texttt{scikit-learn} is integrated with a library for SVM (LIBSVM)\footnote{\url{https://www.csie.ntu.edu.tw/~cjlin/libsvm/}}. The hyperparameter and the tuning steps for SVM are essentially the same as for logistic regression.

Based on the analyses from the coarse parameter space, an exhaustive fine grid search and a randomised search are conducted for each algorithm. The range of values explored are presented in \cref{tab:paramspace}. The best hyperparameter values found for the given classifiers are also provided. Generally, both grid and randomised searches for almost all algorithms yield consistent best estimators, except decision tree.

\begin{table*}
\centering
\caption{Exploration of fine parameter space by the algorithms.}
\label{tab:paramspace}
  \setlength{\tabcolsep}{3pt}
  \begin{threeparttable}
  \begin{tabular}{@{\extracolsep{4pt}}l l l c l c@{}}
    \toprule
    Algorithm & \multicolumn{1}{c}{Hyperparameter} & \multicolumn{2}{c}{Grid Search} & \multicolumn{2}{c}{Randomised Search} \\
    \cline{3-4} \cline{5-6}
     & & \multicolumn{1}{c}{Values} & \multicolumn{1}{c}{Best} & \multicolumn{1}{c}{Values} & \multicolumn{1}{c}{Best} \\
    \midrule
    Decision tree & \texttt{max\_depth} & $3 \twodots 15$ & 6 & $3 \twodots 15$ & 5 \\
     & \texttt{max\_features} & $4 \twodots 12$ & 9 & $4 \twodots 12$ & 11 \\
     & \texttt{min\_samples\_leaf} & $50 \twodots 100$ & 74 & $50 \twodots 100$ & 66 \\
    \midrule
    Random forest & \texttt{max\_depth} & $4 \twodots 9$ & 9 & $4 \twodots 9$ & 9 \\
     & \texttt{max\_features} & $4 \twodots 10$ & 7 & $4 \twodots 10$ & 7 \\
     & \texttt{min\_samples\_leaf} & $50 \twodots 90$ & 53 & $50 \twodots 90$ & 54 \\
    \midrule
    Logistic regression & \texttt{C} & $2^{\{-5, -4.96, \ldots, -1\}}$ & \{0.093, 0.120\}\tnote{*} & $2^{\{-5, -1\}}$ & \{0.093, 0.120\}\tnote{*} \\
    \midrule
    SVM & \texttt{C} & $2^{\{-10, -9.95, \ldots, -5\}}$ & 0.002 & $2^{\{-10, -5\}}$ & 0.002 \\
    \bottomrule
  \end{tabular}
  \begin{tablenotes}\footnotesize
    \item[*] Multiple hyperparameter values with the same test scores and are stated within the range.
  \end{tablenotes}
  \end{threeparttable}
\end{table*}

\paragraph{Model Evaluation} \label{para:modelevaluation} ~\\
Based on the fine grid and randomised searches, the final best estimator is acquired from the model with the highest validation F1 score. The classes of the test dataset are predicted using the best model.

A quantitative representation of the feature importances can be allocated based on how much they contribute to the prediction during the training process. For a single decision tree, $T$, the importance of variable $X_{j}$ \citep{Breiman+:1984} is given by
\begin{align}
\text{Imp}(X_{j},T)=\sum_{t \in N_{T}} \mathds{1}(X_{j},t) \frac{N_{t}}{N}\Delta I(s,t),
\label{eqn:dtimp}
\end{align}
where $N_{t}/N$ is the proportion of samples at node $t$ and $\Delta I(s,t)$ is the impurity decrease of a split $s$ at $t$ as specified in \cref{eqn:impuritydecrease}. The quantity $\mathds{1}(X_{j},t)$ is an indicator function that equals 1 when node $t$ splits on input variable $X_{j}$ and 0 otherwise.

The equation can be extended to calculate the feature importance trained with random forest algorithm \citep{Breiman:2001} by taking the mean of \cref{eqn:dtimp} from a number of trees, which is also known as the mean decrease impurity. The value is normalised to have a sum of unity. The feature importance for tree-based estimators is computed using Gini index as the impurity function.

For logistic regression and SVM, the contribution of a feature is taken to be the absolute value of its weight or coefficient, $\vec{w}$.

\section{Results} \label{sec:results}

\subsection{1d2s Statistical Tests}

Two-sample A--D and two-tailed K--S tests between BALQ and non-BALQ samples are performed on each parameter independently. The statistics are presented in \cref{tab:adks_stats}. Only those that are highly significant at $<0.1$ per cent are marked by an asterisk, while we also considered those that are $<5$ per cent as significantly different.

\begin{table*}
\centering
\caption{One-dimensional two-sample A--D and K--S tests between BAL and non-BAL quasars.}
\label{tab:adks_stats}
  \begin{threeparttable}
  \begin{tabular}{@{\extracolsep{4pt}}l c c c c@{}}
    \toprule
    Parameter &  A--D Stats &  A--D $\alpha$ [\%] &  K--S Stats &  K--S $p$-value [\%] \\
    \midrule
    \texttt{alphanu} & 32.94 & 0.04\tnote{*} & 0.19 & $3.05 \times 10^{-7}$\tnote{*} \\
    \texttt{civmgii\_ratioew} & 7.97 & 0.04\tnote{*} & 0.12 & 0.09\tnote{*} \\
    \texttt{civmgii\_ratiofwhm} & 7.47 & 0.06\tnote{*} & 0.09 & 2.06 \\
    \texttt{civ\_ratioskew} & 7.06 & 0.08\tnote{*} & 0.11 & 0.21 \\
    \texttt{w(civ)} & 5.67 & 0.21 & 0.10 & 0.43 \\
    \texttt{fw(mgii)} & 3.50 & 1.24 & 0.11 & 0.17 \\
    \texttt{mgii\_ratioskew} & 3.35 & 1.42 & 0.09 & 3.22 \\
    \texttt{z.pca} & 3.15 & 1.69 & 0.09 & 3.04 \\
    \texttt{civmgii\_diffv} & 3.10 & 1.77 & 0.08 & 4.39 \\
    \texttt{fw(civ)} & 1.87 & 5.42 & 0.07 & 13.70 \\
    \texttt{w(mgii)} & 1.08 & 11.60 & 0.07 & 8.99 \\
    \texttt{imag} & -0.10 & 38.76 & 0.07 & 17.18 \\
    \bottomrule
  \end{tabular}
  \begin{tablenotes}[flushleft]\footnotesize\setlength{\labelsep}{0pt}
    \item {\bfseries Note:} The results are sorted in descending order of significance based on A--D significance level, $\alpha$.
    \item[*] Highly significant at $<0.1\%$.
  \end{tablenotes}
  \end{threeparttable}
\end{table*}

Based on the results, both statistical tests tend to be fairly consistent at 5 per cent level. The parameters that are significant at the $<5$ per cent in the A--D test also manifest in the K--S test. The spectral index, $\alpha_{\nu}$, shows high significance level of 0.04 per cent and $p$-value of $\ll 0.01$ per cent. This might imply that the BALQs and non-BALQs are physically different in some way.

The EW of \ion{C}{iv} is significant at $<1$ per cent, while the \ion{Mg}{ii} EW shows no statistical difference between the populations. The ratio of \ion{C}{iv} and \ion{Mg}{ii} EW is even more significantly different with A--D $\alpha$ of 0.04 and K--S $p$-value of $<0.1$ per cent. Using the K--S test, \citet{Gibson+:2009} obtained a comparably high significant difference at $p<0.01$ per cent for the EW of \ion{C}{iv} between the BAL and non-BAL quasars. They applied an additional condition in their sample requiring the BAL minimum outflow velocity, \subtext{v}{min}, to be less than $-10\,000\,\kms$, such that the \ion{C}{iv} absorption trough is clearly separated and less affected by the line emission.

The FWHM ratios between the two emission lines are highly significantly different between the two samples using A--D statistics with $\alpha$ of 0.06 while the K--S $p$-value is $<5$ per cent. The FWHM of \ion{Mg}{ii} is also significant at $<5$ per cent level. Conversely, the FWHM of \ion{C}{iv} are not significantly different between the two samples. This is in agreement with \citet{Gibson+:2009}, who found no difference in the FWHM of \ion{C}{iv} distributions between BALQ and non-BALQ populations using samples from the SDSS DR5Q \citep{Schneider+:2007} catalogue. In our result, there is statistical evidence that the asymmetry ratios of individual emission lines are different for the two populations, though $\alpha$ and $p$-value are slightly higher but still below 5 per cent level for \ion{Mg}{ii} line.

The PCA redshift and the velocity offsets of the high-ionisation \ion{C}{iv} line relative to the low-ionisation \ion{Mg}{ii} line are marginally significant at $<5$ per cent. There appears to be no significant difference in the absolute $i$-band magnitude distribution between the two samples.

\subsection{Predictions with ML Classifiers}

\Cref{tab:predictclass} presents the confusion matrix for the test dataset after the algorithms have been trained using the training dataset with the best estimators. The true and false positive and negative rates are also supplied. The complementary of true positive rate is the false negative rate. Similarly, the complementary of true negative rate is the false positive rate. Objects that are correctly labelled, i.e., the true positives and negatives, are highlighted in bold.

\begin{table*}
\centering
\caption{Prediction results of the best estimators from each algorithm for the test dataset using grid and randomised searches.}
\label{tab:predictclass}
  \setlength{\tabcolsep}{3pt}
  \begin{threeparttable}
  \begin{tabular}{@{\extracolsep{4pt}}l l c c c c c c c c@{}}
    \toprule
     & Algorithm & \multicolumn{2}{c}{Decision Tree} & \multicolumn{2}{c}{Random Forest} & \multicolumn{2}{c}{Logistic Regression\tnote{*}} & \multicolumn{2}{c}{SVM} \\
    \cline{3-4}\cline{5-6}\cline{7-8}\cline{9-10}
     & Class & non-BALQ & BALQ & non-BALQ & BALQ & non-BALQ & BALQ & non-BALQ & BALQ \\
    \midrule
    \multirow{5}{*}{\rotatebox[origin=c]{90}{Grid Search}} & non-BALQ & \bfseries 356 & 34 & \bfseries 386 & 29 & \bfseries 323 & 22 & \bfseries 342 & 24 \\
     & Rate [\%] & \bfseries 73.40 & 48.57 & \bfseries 79.59 & 41.43 & \bfseries 66.60 & 31.43 & \bfseries 70.52 & 34.29 \\
     & BALQ & 129 & \bfseries 36 & 99 & \bfseries 41 & 162 & \bfseries 48 & 143 & \bfseries 46 \\
     & Rate [\%] & 26.60 & \bfseries 51.43 & 20.41 & \bfseries 58.57 & 33.40 & \bfseries 68.57 & 29.48 & \bfseries 65.71 \\
     \cline{3-4}\cline{5-6}\cline{7-8}\cline{9-10}
     & F1 score & \multicolumn{2}{c}{0.306} & \multicolumn{2}{c}{0.391} & \multicolumn{2}{c}{0.343} & \multicolumn{2}{c}{0.355} \\
    \midrule
    \multirow{5}{*}{\rotatebox[origin=c]{90}{\parbox{2cm}{\centering Randomised Search}}} & non-BALQ & \bfseries 362 & 32 & \bfseries 385 & 29 & \bfseries 323 & 22 & \bfseries 342 & 24 \\
     & Rate [\%] & \bfseries 74.64 & 45.71 & \bfseries 79.38 & 41.43 & \bfseries 66.60 & 31.43 & \bfseries 70.52 & 34.29 \\
     & BALQ & 123 & \bfseries 38 & 100 & \bfseries 41 & 162 & \bfseries 48 & 143 & \bfseries 46 \\
     & Rate [\%] & 25.36 & \bfseries 54.29 & 20.62 & \bfseries 58.57 & 33.40 & \bfseries 68.57 & 29.48 & \bfseries 65.71 \\
     \cline{3-4}\cline{5-6}\cline{7-8}\cline{9-10}
     & F1 score & \multicolumn{2}{c}{0.329} & \multicolumn{2}{c}{0.389} & \multicolumn{2}{c}{0.343} & \multicolumn{2}{c}{0.355} \\
    \bottomrule
  \end{tabular}
  \begin{tablenotes}[flushleft]\footnotesize\setlength{\labelsep}{0pt}
    \item {\bfseries Note:} Columns represent the actual non-BALQ and BALQ classes, while rows represent the predicted classes. Rates are the true and false positive and negative rates in percentage. The true positives and negatives are highlighted in bold.
    \item[*] Multiple best estimators are present but only one is shown here.
  \end{tablenotes}
  \end{threeparttable}
\end{table*}

Overall, random forest classifiers seem to have the highest test scores. However, this model might suffer from high variance (overfitting), as will be discussed in \cref{sssec:biasvarml}. Logistic regression and SVM models tend to perform similarly. Single decision tree models have lower test scores compared to the others, which is likely due to their simplistic nature. A small variation in the test scores is also observed in decision tree when using either the fine grid or randomised parameter space exploration methods, with poorer scoring metrics found for a grid search. As for the rest of the algorithms, the outputs are rather consistent between the two strategies. Indeed, they are exactly the same for logistic regression and SVM.

In general, the prediction power of the estimators are mediocre for the given imbalanced dataset. Tree-based estimators appear to place greater weight in capturing non-BALQ sample correctly, while logistic regression and SVM treat both equally. These can be seen by the percentage rate and number of accurately identified objects for the individual algorithms. In fact, logistic regression models assign slightly more emphasis on the minority class.

\subsection{Feature Importance and Weighting}

\begin{figure*}
\centering
  \begin{subfigure}[t]{0.495\textwidth}
  \includegraphics[width=\textwidth]{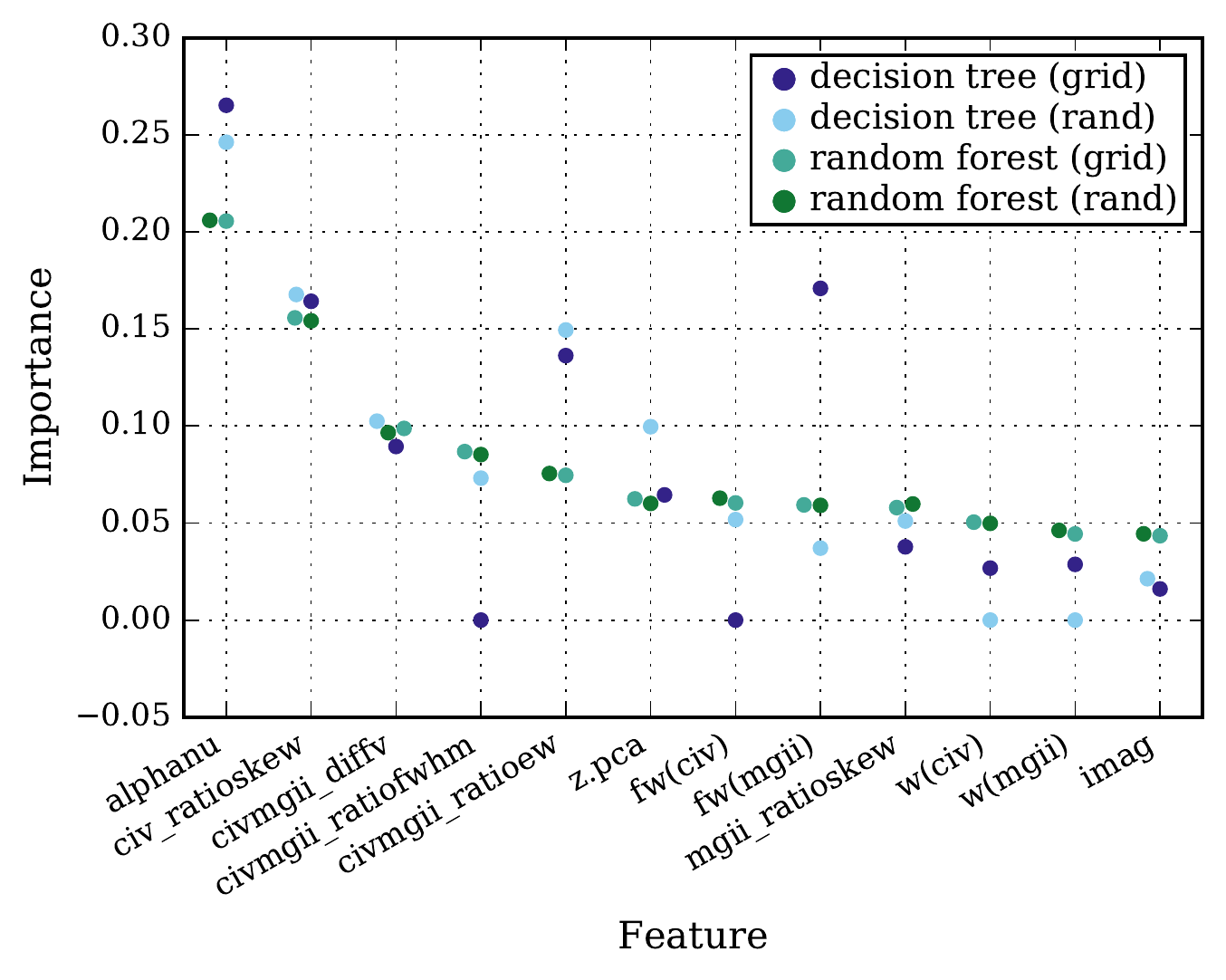}
  \caption{For decision tree and random forest}
  \label{fig:categorical_featureimportance}
  \end{subfigure}%
  \begin{subfigure}[t]{0.495\textwidth}
  \includegraphics[width=\textwidth]{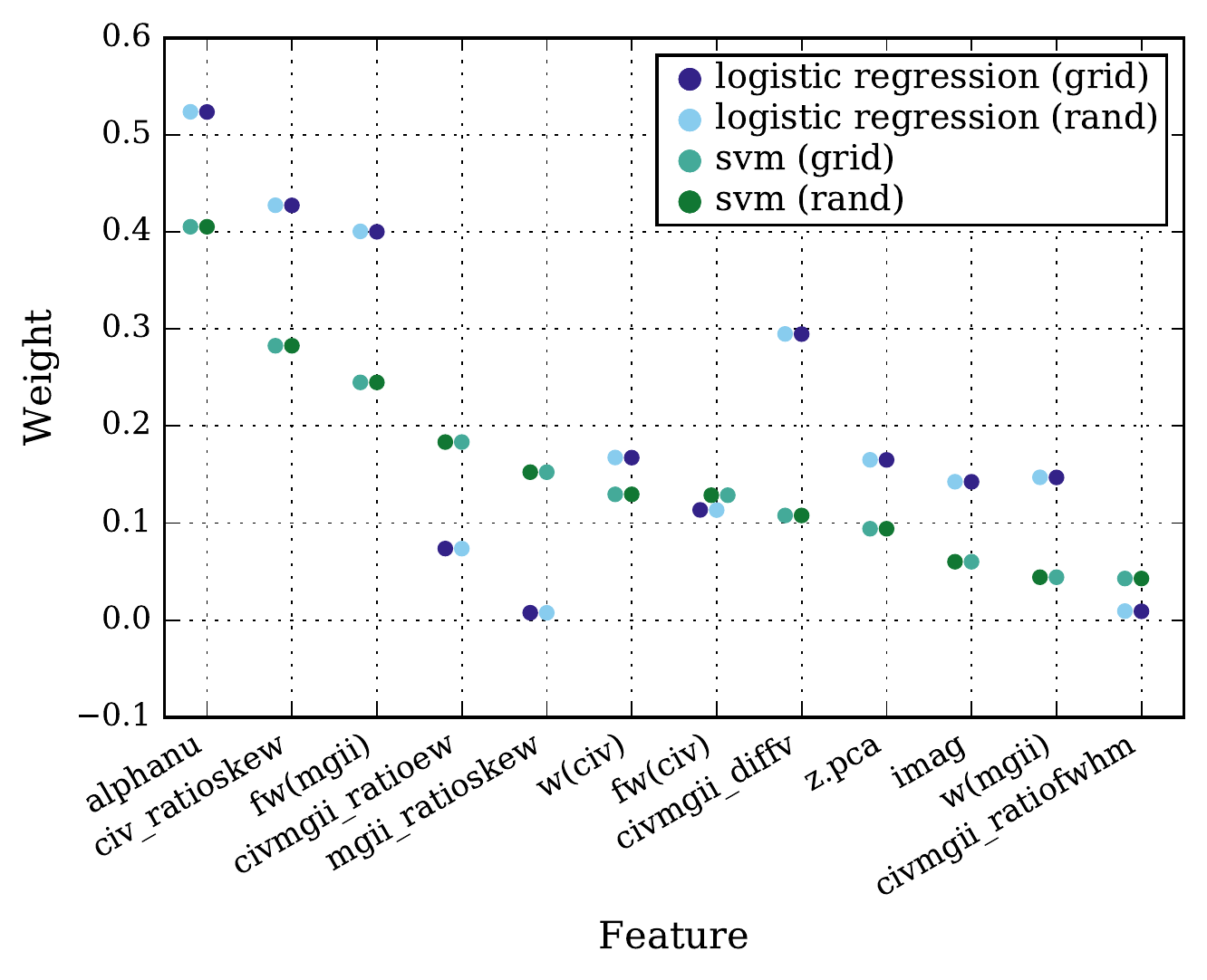}
  \caption{For logistic regression and SVM}
  \label{fig:categorical_featureweight}
  \end{subfigure}
\caption{Feature importance and weighting. Using the best estimator with the highest test scores, the features are sorted in descending order of (\subref{fig:categorical_featureimportance}) importance based on random forest (grid) (\subref{fig:categorical_featureweight}) weighting based on SVM (grid).}
\label{fig:categorical_feature}
\end{figure*}

The feature importance and absolute weighting are computed to identify the contributions of each feature in discriminating the BALQ and non-BALQ groups. The results are depicted in \cref{fig:categorical_feature} and the corresponding values are listed in \cref{tab:categorical_feature}. Higher values correspond to greater deciding factor in the class separation. The three most influential features for individual estimator are highlighted in bold in \cref{tab:categorical_feature}. To compare with the results from 1d2s EDF statistical tests, features that are highly significant at $<0.1$ per cent level and not significant at 5 per cent level are also indicated with symbols.

\begin{table*}
\centering
\caption{Feature importance and weighting for all algorithms}
\label{tab:categorical_feature}
  \setlength{\tabcolsep}{3pt}
  \begin{threeparttable}
  \begin{tabular}{@{\extracolsep{4pt}}l c c c c c c c c@{}}
    \toprule
    Feature & \multicolumn{4}{c}{Importance} & \multicolumn{4}{c}{Weight} \\
     \cline{2-5}\cline{6-9}
     & \multicolumn{2}{c}{Decision Tree} & \multicolumn{2}{c}{Random Forest} & \multicolumn{2}{c}{Logistic Regression} & \multicolumn{2}{c}{SVM} \\
     \cline{2-3}\cline{4-5}\cline{6-7}\cline{8-9}
     & Grid & Rand & Grid & Rand & Grid & Rand & Grid & Rand \\
    \midrule
    \texttt{alphanu}\tnote{*\textdagger} & \bfseries 0.265 & \bfseries 0.246 & \bfseries 0.206 & \bfseries 0.206 & \bfseries 0.524 & \bfseries 0.524 & \bfseries 0.405 & \bfseries 0.405 \\
    \texttt{civ\_ratioskew}\tnote{*} & \bfseries 0.164 & \bfseries 0.168 & \bfseries 0.156 & \bfseries 0.154 & \bfseries 0.427 & \bfseries 0.427 & \bfseries 0.283 & \bfseries 0.283 \\
    \texttt{civmgii\_diffv} & 0.089 & 0.103 & \bfseries 0.099 & \bfseries 0.097 & 0.295 & 0.295 & 0.108 & 0.108 \\
    \texttt{civmgii\_ratiofwhm}\tnote{*} & 0.000 & 0.073 & 0.087 & 0.085 & 0.009 & 0.009 & 0.043 & 0.043 \\
    \texttt{civmgii\_ratioew}\tnote{*\textdagger} & 0.136 & \bfseries 0.149 & 0.075 & 0.076 & 0.074 & 0.074 & 0.183 & 0.183 \\
    \texttt{z.pca} & 0.065 & 0.100 & 0.062 & 0.060 & 0.165 & 0.165 & 0.094 & 0.094 \\
    \texttt{fw(civ)}\tnote{\textdaggerdbl} & 0.000 & 0.052 & 0.060 & 0.063 & 0.114 & 0.113 & 0.129 & 0.129 \\
    \texttt{fw(mgii)} & \bfseries 0.171 & 0.037 & 0.059 & 0.059 & \bfseries 0.400 & \bfseries 0.401 & \bfseries 0.245 & \bfseries 0.245 \\
    \texttt{mgii\_ratioskew} & 0.038 & 0.051 & 0.058 & 0.060 & 0.008 & 0.008 & 0.152 & 0.152 \\
    \texttt{w(civ)} & 0.027 & 0.000 & 0.050 & 0.050 & 0.167 & 0.168 & 0.130 & 0.130 \\
    \texttt{w(mgii)}\tnote{\textdaggerdbl} & 0.029 & 0.000 & 0.044 & 0.046 & 0.147 & 0.147 & 0.044 & 0.044 \\
    \texttt{imag}\tnote{\textdaggerdbl} & 0.016 & 0.021 & 0.043 & 0.044 & 0.143 & 0.143 & 0.060 & 0.060 \\
    \bottomrule
  \end{tabular}
  \begin{tablenotes}[flushleft]\footnotesize\setlength{\labelsep}{0pt}
    \item {\bfseries Note:} The features are sorted in descending order of importance based on random forest (grid), which is the best estimator with the highest test score. The top three highest values are highlighted in bold.
    \item[*] A--D significance level of $<0.1\%$.
    \item[\textdagger] K--S $p$-value of $<0.1\%$.
    \item[\textdaggerdbl] Not significant at $5\%$ level.
  \end{tablenotes}
  \end{threeparttable}
\end{table*}

Notably, every classifier is unanimous in determining the two features in the top three ranking, namely the spectral index and asymmetry of \ion{C}{iv} line. These two features are also highly significant in the A--D tests, while only the spectral index is found to be significant in the K--S test. The FWHM of \ion{Mg}{ii} and the velocity shift between \ion{C}{iv} and \ion{Mg}{ii} that fall into the three most important features in the ML algorithms, are just marginally significant in the statistical test with $p$-value of $<5$ per cent.

Another parameter that is highly significant in both statistical tests is the EW ratio of \ion{C}{iv} and \ion{Mg}{ii} lines. However, it can be seen that this feature is not substantial in splitting the populations with importance values below 0.1 for random forest and logistic regression. Even though the FWHM ratio of the two lines is significantly different between the samples at $<0.1$ per cent in the A--D test, the importance is below 0.1 and even equals zero for decision tree using grid search.

\begin{figure*}
\centering
\includegraphics[width=\textwidth]{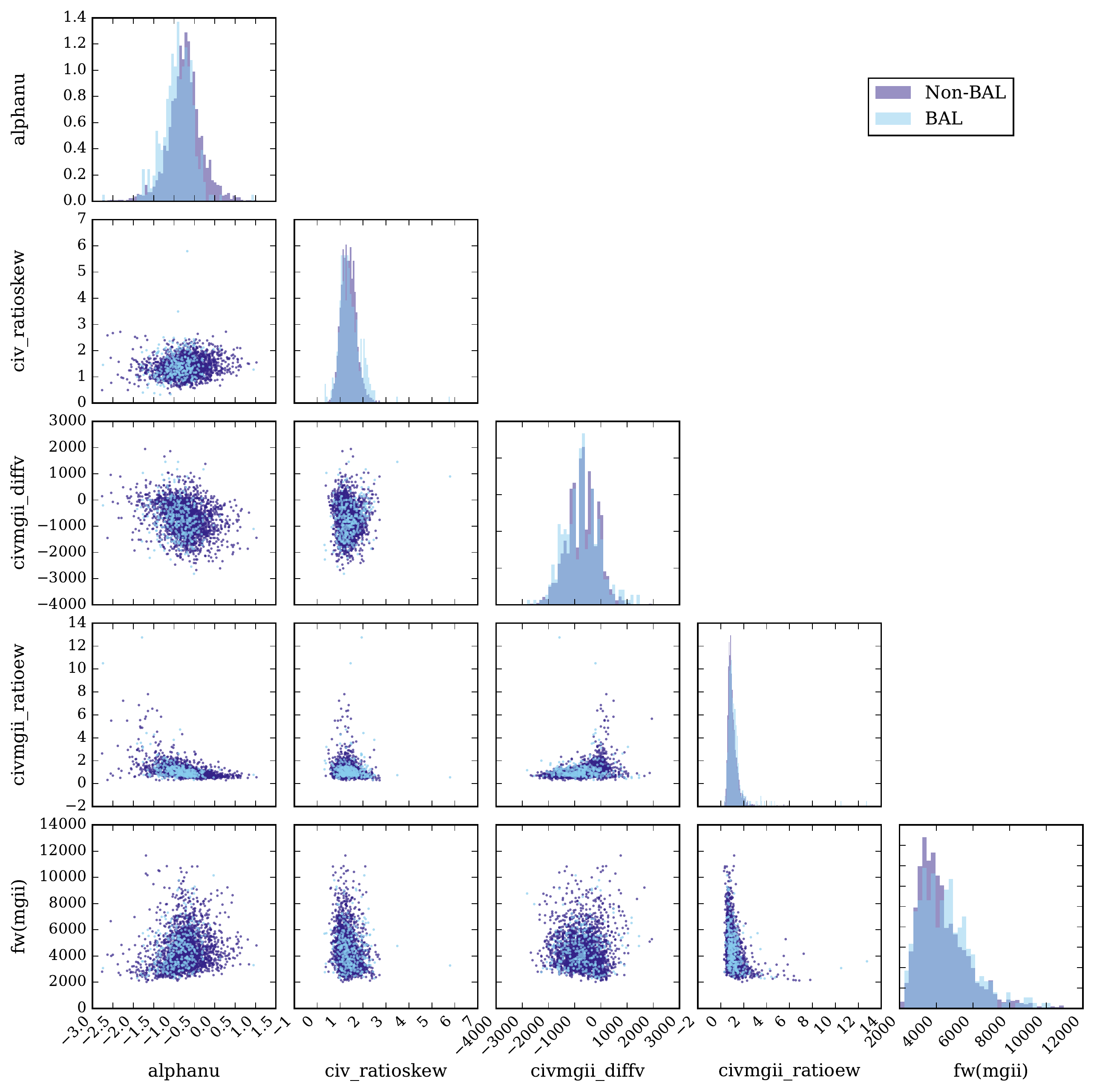}
\caption{Pairwise correlation plot of top three features from each model. The histograms are normalised with the area underneath equals unity.}
\label{fig:pairwisecorr}
\end{figure*}

The distributions for three of the dominant features from each ML estimator are plotted in histograms and scatters as displayed in \cref{fig:pairwisecorr}. The spectral index distribution for the BALQ group appears to be shifted towards lower values with respect to that of non-BALQ group. BALs also show more extreme blue HWHM of \ion{C}{iv} line, though the two histograms are comparatively alike on the lower end. There is a relatively higher number of BALQs with blueshifted \ion{C}{iv} line from \ion{Mg}{ii} line. In principal, the velocity shifts between the two lines are predominantly blueshifted, in agreement with previous studies \citep[e.g.,][]{Gaskell:1982}. The EW ratio of the two BELs for the BALQ sample tends to be marginally larger than for non-BALQ sample. The BALQ population also exhibits a higher median in the \ion{Mg}{ii} FWHM distribution.

In all cases, the shape of the distributions are fairly similar. Based on the scatter plots, there seems to be no obvious trends that can possibly distinguish between the BALQ and non-BALQ classes. Clearly, it is insufficient to isolate the classes using only one or two features. In fact, the similarities in the BALQ and non-BALQ optical/ultraviolet emission line and continuum properties, apart from the excess \ion{N}{v} emission and redder continua in BALQs, are well-established and have been shown in several comparative studies \citep[e.g.,][]{Weymann+:1991,Reichard+:2003}.

The results from the 1d2s EDF tests yield three parameters that are not significant at 5 per cent, which are FWHM of \ion{C}{iv}, EW of \ion{Mg}{ii}, and absolute magnitude in $i$-band. All of these features are also regarded as low importance by the ML classifiers.

Generally, there appears to be some degree of consistency between the different ML classifiers employed. In particular, the features that are ranked prominently such as the spectral index and \ion{C}{iv} line asymmetry by the tree-based estimators, also persist in logistic regression and SVM estimators. Most models are coherent in assigning low values to features that have minimal contribution, such as FWHM ratio of \ion{C}{iv} and \ion{Mg}{ii}, FWHM \ion{C}{iv}, \ion{Mg}{ii} line asymmetry, EW \ion{C}{iv}, EW \ion{Mg}{ii}, and $i$-band absolute magnitude.

\section{Discussion} \label{sec:discussion}

As discussed in the Introduction, there are some pieces of evidence that support the disk-wind model. However, the dynamics and photoionisation structure of the wind are still uncertain. Several aspects of the wind kinematics appear to be well-established. Firstly, gas leaving the accretion disk should retain the angular momentum of the disk in line-driven winds or conserve gas angular velocity if the winds can co-rotate, at least close to the disk, as in magnetocentrifugally driven winds \citep{Proga:2007}. Thus, an outflowing wind would have a helical structure. In addition, there is strong evidence for a dominant outflowing component in at least some parts of the broad emission line region due to the observed blueshift between \ion{C}{iv} and \ion{Mg}{ii} lines. And finally, only the near side of the disk-wind will be seen due to the opacity of the disk. Larger scale emission, such as the narrow line region would be much less obscured.

An outflowing helical wind, viewed along different lines-of-sight will appear to have different measured physical characteristics, for example the FWHM of the BELs. The effect of orientation, particularly for winds with relatively narrow opening angles, is explored in recent papers \citep{Yong+:2016,Yong+:2017}. These papers establish two important results. Firstly, angle of viewing has a significant impact on the measured physical parameters of the BELs. If an emission line is located in a region of the wind with a non-negligible poloidal velocity, then from some angles of viewing, the velocity offset of the line from the systemic velocity will be large. Conversely if the line is emitted from gas that has primarily rotational kinematics, then its peak velocity will largely reflect the systemic velocity of the quasar. Secondly, the FWHM of the line will vary with the angle of viewing, with the exact relationships dependent on the contributions of poloidal, rotational, and wind opening angle to the relevant part of the wind.

Clearly, the similarities and differences in the properties of the emission lines between BALQ and non-BALQ populations provide crucial information on the geometry and physics of the BLR. In a disk-wind model, the common hypothesis is that orientation effects determine whether a BAL or a non-BAL is detected \citep{Murray+:1995,Elvis:2000,Elvis:2004}. In these models, a BAL is predicted to be seen through the narrow outflowing wind, along the line-of-sight of the observer. Though other quasar parameters, such as the black hole mass and mass accretion rate, might be important, they do not mitigate the effect of the AGN's orientation on the line shape. The scatter from the black hole mass and accretion rate will essentially contribute to the spread in the width of the distributions of BEL properties. However, the underlying trends will remain the same, and therefore differences between them can be differentiated by statistical tests. As noted in \citet{Yong+:2017}, varying the black hole mass mainly affects the line width, as predicted from the virial motion, while the degree of ionisation is affected by the mass accretion rate. Generally, the trends in the line profiles, like blueshifting and line asymmetry, still persist. Hence, in a representative population of quasars, BALQs and non-BALQs will appear significantly different in a narrow disk-wind model, as viewed from varying range of angle, despite the expected variation in black hole masses and accretion rate. The differences in the emission line shapes should also be indicative of the structure of the associated wind region. Therefore, any observed difference between the two samples should enable us to deduce the spatial origin of a particular emission or absorption line.

\subsection{Orientation vs.\ Evolutionary Models} \label{ssec:ovse}

It remains unclear from the literature which model, the orientation or evolutionary, can provide a definite explanation on the presence of BALs or whether a combination of both model is required. When considering the orientation model, most authors have assumed the BEL disk-wind to have a structure similar to \citet{Elvis:2004}, with a fairly narrow wind opening angle. The similarity between the two samples in this analysis is in conflict with this model. Also, models that invoke a narrow wind, do not all agree on the orientation of the wind. In fact, there are several pieces of evidence that appear inconsistent with regards to the launching angle of the wind. Evidence of an equatorial wind launched from the accretion disk are supported by spectropolarimetry observations \citep{Goodrich+Miller:1995,Cohen+:1995,Ogle+:1999,Lamy+Hutsemekers:2004,Brotherton+:2006,Young+:2007}. Detailed models incorporating both hydrodynamical wind and photoionisation from radiation field also favour this idea \citep{Proga+:2000,Proga+Kallman:2004}. However, this is complicated by the discovery of radio-loud BALQs with polar outflows, implying that they are seen near parallel to the relativistic radio jet \citep{Zhou+:2006,Ghosh+Punsly:2007}. This hints that BAL outflows can span a broad range of viewing angle.

Suppose the sole factor that governs the intrinsic BALQ fraction is the viewing angle along the narrow wind opening angle. Then, it is predicted that the fraction should remain the same over time, and therefore be redshift independent. However, \citet{Allen+:2011} found that the intrinsic fraction of \ion{C}{iv} BALQs changes with redshift, which they argue contradicts the orientation only explanation. Additionally, \citet{Becker+:2000,Montenegro-Montes+:2008,Fine+:2011} demonstrated that the radio spectral index distributions in a small sample of BAL and non-BAL quasars are comparable, indicating no preference in orientation. Using a larger sample consisting of 74 BALQs, \citet{DiPompeo+:2011} discovered that there are more radio core sources in BALQs, associated with a face-on inclination, but with both BALQ and non-BALQ populations having the same spectral index range. \citet{DiPompeo+:2012b} also found no relationship between the radio spectral index and BAL outflow properties. Each of these lines of evidence, argue against a limited angle of viewing for BALQs.

There are also some inconsistencies between the observations and the predictions of the evolutionary model. In the optical/ultraviolet wavelength regime, BALQs exhibit higher reddening than non-BALQs, which could indicate that a quasar with BAL is younger and dwells in a dustier environment \citep[e.g.,][]{Weymann+:1991,Sprayberry+Foltz:1992}. Our results in the previous section also show that there is a distinct difference in the spectral indices between BALQ and non-BALQ samples, which might be due to dust extinction. The distribution of the spectral index for BALQ sample tends to be smaller than that of non-BALQ sample. This difference in spectral indices between the two populations is also detected by the machine learning algorithms, which yields the spectral index as the dominant feature that distinguish the two groups. A quasar with a higher covering fraction, as deduced in the evolutionary model, is expected to produce more emission in the longer wavelengths, where less dust is being absorbed. However, studies in the mid-infrared \citep{Gallagher+:2007}, far-infrared \citep{CaoOrjales+:2012}, and submillimetre \citep{Willott+:2003} show that the spectral energy distributions (SEDs) are the same for a small sample of BAL and non-BAL quasars with similar luminosity. When using a larger sample of 72 radio-loud BAL objects from \citet{DiPompeo+:2011}, \citet{DiPompeo+:2013} discovered that there is indeed more radiation in the mid-infrared in the BALQ sample. Hence, a mixture of both the orientation and evolutionary models has been suggested \citep{Gallagher+:2007,Allen+:2011,DiPompeo+:2013}. \citet{Lawther+:2018} found close companion galaxies to a subset of observed FeLoBALs supporting a merger-induced young quasar scenario.

If the orientation model were valid, the observed emission line properties of the BAL and non-BAL quasars are predicted to be distinct in a narrow wind disk-wind model. We can rule out BALs arising from specific fixed narrow wind opening angle if the BEL properties are statistically indistinguishable. This is because the resulting BALQs and non-BALQs will populate different extremes of the BEL property distributions depending on the angle the disk wind is located. For BALs viewed through a polar narrow wind, the line profile will be narrower and more blueshifted compared to those for non-BALQs that are seen at non-polar angles. The opposite is true for equatorial BAL wind. Additionally, if the narrow wind lay in the middle of these extremes, we expect that the non-BALQs to show both the most narrow, blueshifted and the most wide, symmetric BEL profiles, while the BALQs would not. This would manifest as a difference between the BEL properties of BALQs and non-BALQs, and can be identified by statistical tests. For a rotationally dominated structure, the FWHMs should increase significantly with inclination angles toward edge-on, as a higher portion of the velocity aligns with the line-of-sight \citep{Yong+:2017}. In this case, a systemic shift to larger FWHM values should be detectable by the statistical tests and machine learning. If the evolutionary model were valid, we could expect that there would be no difference in most of the emission line features for the two groups, assuming the only difference between the two groups is the shrouding of the AGN by a dust and gas cocoon. However, it is possible that we are looking at very different AGN states due to different accretion rates and less relaxed systems, and hence differences in emission properties may arise.

In our analysis, the BAL and non-BAL characteristics are found to be very much alike, which argues against the idea that a BAL is formed when the projected line-of-sight intercepts an equatorial outflowing wind with narrow opening angle. Although we assume that the BELs are formed in the BAL wind, our argument holds as long as the BEL emitting region has a realistically flattened axisymmetric geometry. In fact, only a few attributes of the BELs are highly significantly different at $<0.1$ per cent level. The outcomes from ML also support the difficulty in discriminating the BALQs from the non-BALQs. We found that with only the observed continuum and BEL features, the two populations cannot be clearly distinguished. Even the features that demonstrated significant differences still show strong similarities in the overall distributions. This suggests that a revision in the commonly used paradigms is required to describe the observed trends.

\subsection{Caveats} \label{ssec:caveats}

\subsubsection{Dataset}

It is noteworthy to mention a few caveats that would affect our results. As datasets are normally heterogenous, they are prone to suffer from selection bias. This might shift the significance of the results and the observed trends, but it should not substantially alter our major conclusion.

The true BALQ fraction is still poorly established. In our sample, the percentage of quasars with BALs is $\sim 11.3$ per cent within redshift of $1.57 \leq z \leq 2.42$, which is slightly lower than other published results. Variation in the observed fraction have been found to be $15 \pm 3$ per cent for $1.5 \leq z \leq 3.0$ \citep{Hewett+Foltz:2003}, $14.0 \pm 1.0$ per cent for $1.7 \leq z \leq 4.2$ \citep{Reichard+:2003}, $13.3 \pm 0.6$ per cent for $1.68 \leq 2.3$ \citep{Gibson+:2009}, and $8.0 \pm 0.1$ per cent for $1.6 \leq z \leq 5$ \citep{Allen+:2011}. After taking into account certain selection effects, the intrinsic fraction of BALQs are generally found to be larger, for example, about $22 \pm 4$ per cent \citep{Hewett+Foltz:2003}, $15.9 \pm 1.4$ per cent for $1.70 \leq z \leq 3.45$ \citep{Reichard+:2003}, $16.4 \pm 0.6$ per cent \citep{Gibson+:2009}, and $40.7 \pm 5.4$ per cent \citep{Allen+:2011}.

The number of BALQs in any given sample is dependent on the choice of constraints applied to classify BALQs and non-BALQs. Different S/N cuts affect the BALQ detection rate. There is a higher chance of identifying a BALQ in spectra with high S/N \citep{Allen+:2011}. In addition, the BI measure used to determine a BAL might underestimate the total BALQs as it rejects those with weak absorption troughs.

The measure of absorption trough relies on the method of line fitting applied on the quasar samples \citep{Paris+:2017}. The emission lines in SDSS DR12Q are fitted using PCA components created using a sample of non-BALQ spectra. Each line is fitted with five PCA components and the continuum is fitted as a power-law using the best fit PCA component around the emission line. \citet{Paris+:2017} reported that the PCA method produces a similar BI distribution to that of \citet{Allen+:2011} using non-negative matrix factorisation, but differs slightly from \citet{Gibson+:2009}, which shows more objects with low BI values. \citet{Gibson+:2009} modelled the emission lines using a power-law continuum and a Voigt profile. Therefore, some BALQs may have been misidentified as non-BALQs or vice versa.

We did not inspect every spectra on the accuracy of the each parameter value extracted from SDSS DR12Q \citep{Paris+:2017} catalogue. It is possible that some spectra, especially the BALQ sample, are not fitted accurately since the fitting procedure is done automatically and is based on a sample without broad absorption features. Consequently, the results could be biased towards the properties of the non-BALQs. Additionally, the measurement uncertainties for each property are not accounted for.

\subsubsection{Bias-variance Trade-off in Machine Learning Algorithms} \label{sssec:biasvarml}

To determine if increasing the size of the training dataset will help to boost the score, we examine the learning curves with stratified 10-folds CV of every algorithm. The learning curve also serves to identify situations where ML models suffer from under or overfitting.

It is found that the properties used could not provide a good fit, in the sense that the training and validation scores are low no matter the size of the training sample. Using a more complex ML model, such as multi-layer perceptron and incorporating boosting methods, might marginally increase the scores. However, given the input features, the prediction capabilities using different ML classifiers is not expected to significantly boost the test scores. This is shown by the fairly similar test scores of the four different ML algorithms. An alternative method of improving the classification algorithm is to add more features. However, this requires some insight into which features are actually substantial in differentiating the BALQ from the non-BALQ group, which is not obvious at present.

Additionally, random forest algorithms show signs of high variance, with a huge gap between both scores. A slight indication of this problem is also seen in decision tree algorithms. This means that the training dataset is overfit, and therefore gives poorer results when the algorithm is applied to the test set. To counteract this dilemma, increasing the training samples and reducing the complexity of the model might remove this gap. In contrast, logistic regression and linear SVM are free from this issue. The learning curves of both scores have somewhat converged. In this case, adding data will not further improve the results.

\section{Proposed BLR Disk-wind Model} \label{sec:proposedblr}

In general, there is a large diversity in line widths and line shifts between the different BELs in a single object. When this information is combined with the line strengths this variety can be utilised to infer the ionisation potential and dynamics of the gas. The different properties of the high-ionisation \ion{C}{iv} line and low-ionisation \ion{Mg}{ii} line suggest these two lines are emitted in distinct regions.

It is still unknown whether BELs and BALs originate from the same region or structure, or the nature of this region. As mentioned previously, some of the observed trends are inconsistent with the published disk-wind models, incorporating a narrow outflowing wind angled close to equatorial and emanating from an opaque accretion disk. This hints that other BLR outflow geometries are needed to understand the observations.

\begin{figure*}
  \centering
  \includegraphics[width=0.7\linewidth,keepaspectratio]{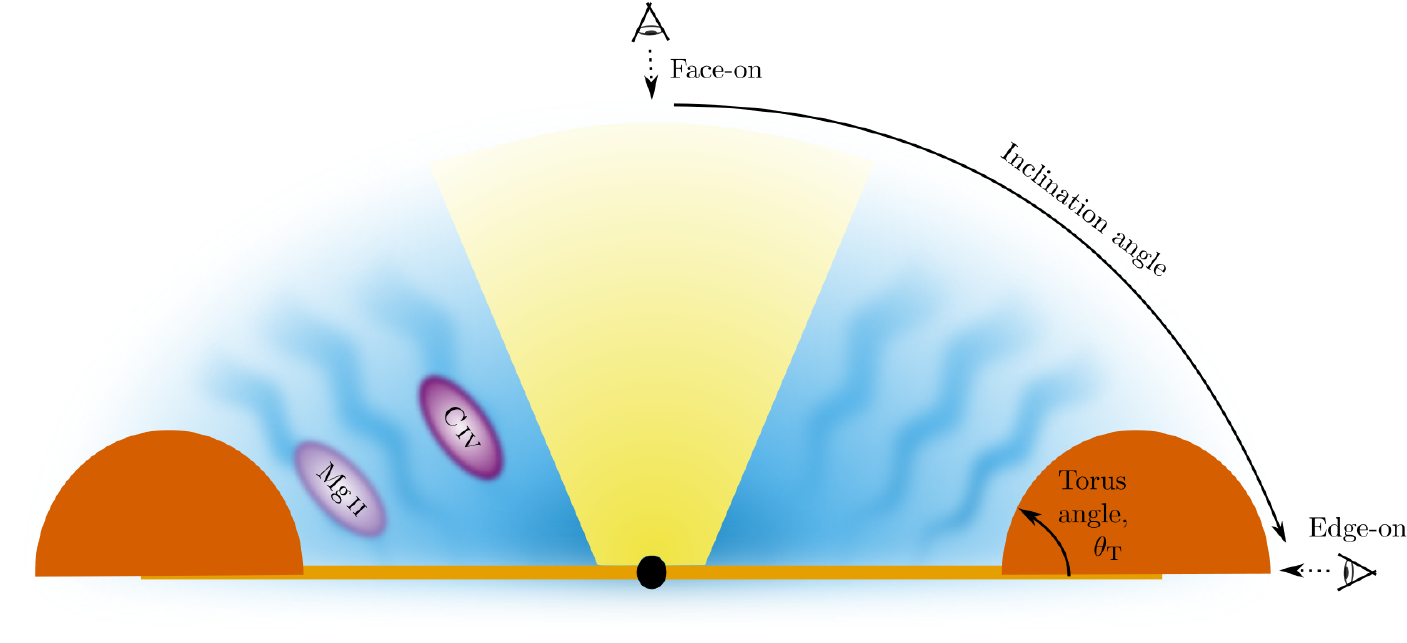}
  \caption{The proposed BLR disk-wind model with the following features: black hole (black), optically thick accretion disk (orange), dusty torus (vermillion), ionisation cone (yellow), and wind that spans a wide range of angles (sky blue). The wind is driven by some acceleration mechanism, forming radial streams and are regions with denser gas (curl lines in the wind). These regions can be transient in nature and occur at different angles in each quasar. BALs are seen when the line-of-sight intersects these regions. The bulk of the high-ionisation \ion{C}{iv} line emission (dark purple) is emitted close to the black hole, while the bulk of the low-ionisation \ion{Mg}{ii} line emission (light purple) is emitted near the base of the wind and further from the centre of the ionising source. The region in the ionisation cone is completely ionised. The view of the inner regions will be obstructed by the dusty torus when seen with an angle \subtext{\theta}{T} of edge-on.}
  \label{fig:blrdwmodel}
\end{figure*}

To address this situation, we present a potential disk-wind model of BLR with the assumption that the BLR is directly associated with the BAL wind. The highlights of our model include the following important additional features:
\begin{itemize}[leftmargin=.125in,listparindent=\parindent,itemsep=1ex]
  \item The wind spans a wide range of angles between the torus and an axial ionised cone.
  \item Within that wind, there are multiple radial streams of higher density, clumpy material.
  \item BAL troughs are detected if an observer's line-of-sight intersects any of these streams, while non-BALQs are seen for other lines of sight through the wind.
\end{itemize}
The model is illustrated in Fig.~\ref{fig:blrdwmodel} and the specifications of the features are elaborated as follows:
\begin{itemize}[leftmargin=.125in,listparindent=\parindent,itemsep=1ex]
  \item \textbf{The BLR comprises a disk-wind with a rotational velocity component reflecting the angular momentum of its origin in the accretion disk, and a poloidal component resulting from an acceleration mechanism. The emission line flux produced at each region of the wind is expected to reflect the local density and ionisation state.}\\
  The main acceleration mechanism is not known but could be due to a gradient in gas pressure or thermal expansion \citep{Weymann+:1982,Begelman+:1991}, radiation pressure or line driving \citep{Shlosman+:1985,Arav+:1994,Murray+:1995}, or it could be a magnetically driven wind \citep{Blandford+Payne:1982,Emmering+:1992,Konigl+Kartje:1994}. The presence of the BALs strongly indicates that substantial momentum is transferred from a powerful radiation field to the gas. The preferred driving mechanisms for a disk-wind are often a combination of line-driven and magnetic field \citep{Konigl+Kartje:1994,deKool+Begelman:1995,Proga:2003,Everett:2005}.
  
  Our understanding on how line driving can produce powerful, high velocity winds is based on studies of winds in hot stars \citep{Castor+:1975}. Luminous and massive hot stars possess fast winds that are radiatively line-driven due to the intense radiation of the star \citep{Castor+:1975,Pauldrach+:1986,Friend+Abbott:1986}. Subsequently, line driving mechanisms has been adapted in other accreting systems, such as cataclysmic variables, quasars and young stellar objects. Hydrodynamical models of AGN using line driving have successfully created winds of the observed dynamics \citep{Proga+:2000,Proga+Kallman:2004}. However, the efficiency of line driving is highly dependent on the ionisation state of the outflow. Due to the strong X-ray emission of AGN, without a layer of `hitchhiking' gas \citep{Murray+:1995} or a small filling factor (we will discuss this further), the gas will be too ionised to drive the wind.
  
  Further evidence for the role of radiation in driving the wind is the blueshift of emission lines. \citet{Richards+:2011} conducted a comprehensive investigation on the correlation of \ion{C}{iv} blueshift with quasar properties and showed that the trends can be explained within the framework of disk and wind components. They suggested that weak \ion{C}{iv} with large blueshift is indicative of a wind-dominated system with relatively low X-ray luminosity that allows strong winds to be driven via radiative pressure. In a disk-dominated system, the relatively high X-rays will overionise the wind and suppress line-driving mechanism. \citet{Shen+:2016} found that the blueshift of \ion{He}{ii}, \ion{C}{iv}, and \ion{Si}{iv} have strong luminosity dependence. Their average blueshift relative to LILs such as [\ion{O}{ii}] or \ion{Mg}{ii} increases with luminosity, consistent with previous findings \citep[e.g.,][]{Hewett+Wild:2010,Richards+:2011,Shen+:2011,Shen+Liu:2012}.
  
  We expect that magnetic fields may be important in driving or confining the ionised outflowing wind. Magnetic fields are essential to the existence and evolution of accretion disks as magnetorotational instability is almost certainly responsible for local angular momentum transport in accretion disks \citep{Balbus+Hawley:1991}. Therefore, it is likely that the magnetic fields play a role in the wind. Magnetocentrifugal wind models are able to predict the geometry and kinematics of a wind, but considerable assumptions must be made on the mass-loss rate, making a magnetocentrifugal wind less robust to observational testing.

  \item \textbf{Only the near side of the BLR disk-wind is observed due to the opacity of the accretion disk.}\\
  This is a commonly accepted argument due to the high column density of the accretion disk. However, the assumption does depend on the opacity remaining high out beyond the self gravitation radius.

  \item \textbf{A region or cone around the accretion axis may be completely ionised, and therefore not contribute to the BELs.}\\
  Due to the strong X-ray component of the AGN emission, the gas close to the central source is expected to be overionised. \citet{Proga+:2000,Proga+Kallman:2004} found a region consistent with this in their line driving simulations.

  \item \textbf{The BLR will be obscured for viewing angles close to edge on (i.e., the plane of the accretion axis). Thus, the BLR will be observed in AGN viewed from a range of angles, significantly less than a solid angle covering of $4\pi$.}\\
  This assumption is based on the general unification model of AGN, where the obscuration comes from a type of dusty torus, thickened accretion disk, or dusty outflow \citep[e.g.,][]{Antonucci:1993}. The existence of dust on scales just beyond the BLR is well established \citep{Jaffe+:2004,Kishimoto+:2009,Kishimoto+:2011,Honig+:2012,Kishimoto+:2013}, and can explain the observed broad line emission in the polarised spectra of some type 2 quasars \citep[e.g.,][]{Antonucci+Miller:1985,Miller+Goodrich:1990,Tran+:1992,Heisler+:1997,Zakamska+:2005}.

  \item \textbf{Within the wind, the ionisation (and density) will vary depending on the angle from the accretion axis and radial distance from central source. In particular, HILs such as \ion{C}{iv} will be preferentially emitted closer to the accretion axis, while LILs such as \ion{Mg}{ii} and H$\alpha$ will be emitted from regions closer to the accretion disk.}\\
  This assumption is based on the line profiles of the associated lines. LILs tend to be more symmetric, having little or no velocity shift from the systematic redshift of the object. Their line widths are consistent with their motions being dominated by Keplerian motion, and hence are often used to estimate the black hole masses \citep{McLure+Jarvis:2002,Shen+:2008,Wang+:2009,Rafiee+Hall:2011,Trakhtenbrot+Netzer:2012,Mejia-Restrepo+:2016}. In contrast, HILs are commonly blueshifted with respect to the systematic redshift and show significant asymmetry. Shape corrections are required to recover black hole masses consistent with those found with LILs \citep{Collin+:2006,Shen+Liu:2012,Denney:2012,Park+:2013,Runnoe+:2013b,Yong+:2016,Coatman+:2017,Park+:2017}. This may suggest another velocity component is playing a significant role in the motion of the HIL emitting gas. The models of \citet{Yong+:2017} show that emission lines emitted from low in a disk-wind are more symmetric and unshifted from the systematic redshift, while lines emitted from regions close to the central source with a larger angular offset from the disk show significant blueshifts for most inclinations, and have a substantial poloidal velocity component causing line asymmetries.

  \item \textbf{The BAL wind is co-spatial with the BLR or at least part of the same extended geometry.}\\
  As the BELs are often absorbed, the BAL wind must be either outside or co-spatial with the BLR. We expect that the BEL and BAL region originate from the same continuous geometry in the BLR, driven by same mechanism, and not distinct regions. This is due to the similar ionisation states required to emit and absorb the observed lines. Additionally, some connections have been found between BALs and BELs, such as correlations between the minimum outflow velocity (detachment) of the BAL troughs and the width or the strength of the BELs \citep{Turnshek:1984,Lee+Turnshek:1995} and the BEL blueshifts are particularly large in BALQ composite spectra \citep{Richards+:2002}.

  \item \textbf{BALs will be observed when the angle of viewing intersects a region of denser gas. This suggest that these regions occur throughout the wind, not in a preferred direction.}\\
  The similarity between the BAL and non-BAL quasar FWHM distributions suggest that there is not a significant inclination difference between the two samples, as discussed in \cref{ssec:ovse}. This means that BALQs are viewed from similar angles to non-BALQs and a flattened or narrow disk-wind model, like that proposed by \citet{Murray+:1995,Elvis:2000,Elvis:2004}, are inconsistent with our finding. Instead we propose a wind covering a large angular range that is intrinsically clumpy in nature. These clumps are embedded along streams and are regions of high column density. Multiple radial streams may be present, presumably with an integrated covering factor of $<20$ per cent to account for the observed fraction of BALQs to non-BALQs (assuming a non-evolutionary argument). BALs are observed when our line-of-sight passes through these streams. The BELs are also produced. The regions are radially extended, spanning the wide velocity range observed in BAL troughs, whereas they can have orientations anywhere in the viewing angle range of non-BALQs. They are temporary with differing launching angles in individual quasars as this randomness works well with the observed timescale of variability in the BALs \citep{Hamann+:2013}. These regions can be thought of as radial streams of clumps with a range of optical depths. While they are expected to be mostly optically thick, overall these regions will have different density and optical thickness that give rise to the distinct elements in the spectrum. The scale of these clumps is estimated to have lower and upper limit radial outflow sizes of $10^{-7}\text{--}10^{-4}\,$pc \citep{Hamann+:2013,McGraw+:2017} with transverse outflow sizes of $\gtrsim 2.6 \times 10^{-3}\,$pc \citep{Hamann+:2013}.
  
  The idea of small-scale dense substructure has already been well established in hot stars. Line-driven winds in stellar winds can be unstable and are affected by line-deshadowing instability \citep[LDI;][]{Lucy+Solomon:1970,MacGregor+:1979,Abbott:1980,Carlberg:1980,Lucy+White:1980,Owocki+Rybicki:1984}. Time dependent numerical simulations have shown that LDI is able to create strong wind shocks, which leads to the compression of materials in the star atmosphere and the formation of a highly inhomogeneous dense clumped structure \citep[e.g.,][]{Owocki+1988,Feldmeier+:1997}. The existence of clumped winds in massive stars is also confirmed by observations \citep[see reviews by e.g.,][]{Hamann+:2008,Sundqvist+:2012}. Additionally, the numerical studies on cataclysmic variable stars also demonstrate that the clumps in the line-driven winds are time dependent \citep{Proga+:1998,Dyda+Proga:2018}.
  
  In AGN models, a clumpy wind is naturally produced in both a magnetocentrifugal outflowing wind \citep{Emmering+:1992} and a line-driven wind \citep{Proga+Kallman:2004}, or may represent transient density enhancements formed due to turbulence or shocks in radiatively driven wind \citep{Arav+:1994}. A clumpy wind is also one of the two main mechanisms proposed to resolved the overionisation problem found for all wind models regardless of driving mechanism. A sufficiently low filling factor increases the electron number density and lowers the ionisation parameter enough to prevent overionisation \citep[e.g.,][]{deKool:1997,Hamann+:2013,Baskin+:2014}.
  
  \citet{Sim+:2010} and \citet{Higginbottom+:2013,Higginbottom+:2014} performed a Monte Carlo radiative transfer simulation of line-driven disk-wind and managed to produce synthetic spectra of AGN. However, they noted that the wind in high luminosity X-ray sources tends to overionised and consequently inhibits the production of ultraviolet absorption lines. An extension to their work by \citet{Matthews+:2016} found that a clumpy wind was required to moderate the ionisation state of the gas and allow the formation of BAL features at realistic X-ray luminosities. It allowed the formation of strong emission lines, although their simulations were still not able to recreate all the emission lines seen in quasar spectra with the correct EW ratios. \citet{Everett+:2002} also found that a clumpy, multi-phase outflowing gas could explain the observed BAL spectra well, and models without clumping fail to explain how different ionisation absorption lines are found to have similar velocity structures.
  
  There is also further evidence of dense substructures in AGN winds from BALQ absorption line profiles. BALQs show complex absorption line profiles, commonly consisting of a number of distinct troughs, each only a few 1000\,\kms\ broad \citep[e.g.,][]{Korista+:1992,Hamann:1998,Arav+:2001,Trump+:2006,Ganguly+:2006,Gibson+:2009,Simon+Hamann:2010} and exhibit variability in these profile shapes \citep[e.g.,][]{Barlow:1994,Gibson+:2008,Capellupo+:2011,Capellupo+:2012,Capellupo+:2013,Grier+:2015}, which suggests motion of absorbing gas transverse to our line-of-sight \citep{Hamann+:2013}. In addition, BALs display a wide range of ionisation levels, which is inconsistent with a single uniform-density absorber \citep{Turnshek+:1996,Hamann:1997}.

  \item \textbf{The covering fraction of the clumps and streams of dense clumps may vary as a function of age of the quasar or accretion rate.}\\
  In this case, the probability of seeing BALs in any particular quasar is still a function of the covering fraction of the absorbing material, but the coverage itself is a function of time or accretion rate. This helps explain the observed redshift evolution of the two populations found in \citet{Allen+:2011} and could be related to the difference in spectral index that we see.
\end{itemize}

We also attempt to explain some aspects of the observed phenomena:
\begin{itemize}[leftmargin=.125in,listparindent=\parindent,itemsep=1ex]
  \item \textbf{LoBALs are HiBALs.}\\
  Quasars with LoBALs also have HiBALs, but not the opposite. Within our model, the presence of LoBALs and FeLoBALs is due to the enhanced covering fraction, enhanced Hydrogen column density, and to a lesser degree a smaller ionisation parameter. The generally weak [\ion{O}{iii}] emission and strong reddening of LoBALs suggests that LoBAL quasars tend to be surrounded by dust and gas that has a larger global covering factor compared to HiBALs \citep[e.g.,][]{Boroson+Meyers:1992,Turnshek+:1994,Zhang+:2010}. Additionally, \citet{Liu+:2015} found using photoionisation modelling that the column density of the respective ion species, \subtext{N}{ion}, and thus the absorption strength depends on the ionisation parameter, $U$, and the cloud/outflow thickness (described by the cloud's Hydrogen column density, \subtext{N}{H}). They found that the \ion{C}{iv} column density, \subtext{N}{\ion{C}{iv}}, was several orders of magnitude larger than that of \ion{Mg}{ii} at small \subtext{N}{H}, but \subtext{N}{\ion{Mg}{ii}} increased significantly to become comparable with \ion{C}{iv} at a sufficiently large \subtext{N}{H}. \subtext{N}{\ion{Mg}{ii}} is also dependent on $U$ and a larger \subtext{N}{H} is required for a system with a larger $U$ to obtain the same \subtext{N}{\ion{Mg}{ii}} value. If the cloud's \subtext{N}{H} is sufficiently large, both high- and low-ionisation absorption lines will be detected. \citet{Baskin+:2014} also found LoBALs present in high \subtext{N}{H} radiation pressure confined gas slab, and that the corresponding absorption of the higher ionisation lines (e.g., \ion{C}{iv}) was significantly more complete.
  
  In addition, LoBAL quasars have been found to display typically larger HiBAL BI values compared to systems with only HiBALs \citep{Allen+:2011}. A higher BI value plausibly indicates that the range of velocities that are absorbed and the absorption depth are large. In the context of our model, this phenomenon is likely to occur when the absorption happens over a large region that contains a myriad of outflow velocities. The gas will have a larger covering fraction, and hence more clumps intercept with our line-of-sight, building up the column density and LoBAL quasars can be detected.

  \item \textbf{The trend between FWHM and EW of \ion{C}{iv} and \ion{Mg}{ii} lines.}\\
  The relationship between the FWHM and EW for the different quasar populations should then reveal details on the structure, kinematics, and dynamics of the BLR. There is an anti-correlation between FWHM and EW for HIL \ion{C}{iv} \citep{Francis+:1992,Wills+:1993,Brotherton+:1994}, while a positive correlation is identified for LIL \ion{Mg}{ii} \citep{Brotherton+:1994,Puchnarewicz+:1997}.
  
  To explain the trend seen in HILs, \citet{Francis+:1992,Wills+:1993} proposed that the BLR comprise a two line-emitting region: a very broad line region (VBLR) and an intermediate line region (ILR). The components of a BEL include a line core and broad line wings, which are emitted from the VBLR and ILR respectively. Since the spherical VBLR is located closer to the central ionising source, it has a higher velocity and density compared to the disk-like ILR. With increasing line core relative to the line wings of the HIL, the EW increases while the FWHM decreases since the peak is narrower and sharper. Meanwhile, the LIL has relatively weak core but strong line wings, which implies that it might originate within the ILR where the velocity is lower \citep{Puchnarewicz+:1997}.
  
  In our model, this trend is explained in combination with the Baldwin effect \citep{Baldwin:1977}. As the continuum luminosity increases and likely the SED softens, the EW of \ion{C}{iv} and \ion{Mg}{ii} decreases. In a lower luminosity quasar, the observed emission is likely to be emitted closer to the central black hole based on the observed radius--luminosity ($R \text{--} L$) relationship \citep[e.g.,][]{Kaspi+:2007,Bentz+:2009}. To explain the trend found for \ion{C}{iv}, we require the wind to be radiation driven or at least the strength of the wind to be positively correlated with the luminosity. In this case, the increase in FWHM for decreasing EW is due to an increase in outflow velocity with increasing luminosity. Additionally, \ion{C}{iv} FWHM is found to correlate with velocity shift, suggesting stronger outflows for these objects. This supports our interpretation. For a line profile dominated by Keplerian motion, expected for \ion{Mg}{ii}, things get more complicated. If a constant black hole mass is assumed, the FWHM should increase as the luminosity decreases with FWHM $\propto L^{-1/4}$, which fits the observed trend. However, the opposite trend is expected if the black hole mass is varied and the relative accretion rate (Eddington fraction) is kept fixed. In this case, the FWHM should decrease as the luminosity decreases with FWHM $\propto L^{1/4}$, that is the FWHM should decrease with increasing EW, assuming that the luminosity is proportional to black hole mass. This is in contradiction to the observed trend. Therefore, the \ion{Mg}{ii} EW vs.\ FWHM trend is not easily resolved in our model using a dynamic driven argument.

  \item \textbf{Variability in the line profile.}\\
  Based on reverberation mapping, it has been found that the response in the red side of most lines is faster than the blue side \citep[e.g.,][]{Gaskell:1988,Koratkar+Gaskell:1989,Crenshaw+Blackwell:1990,Korista+:1995,Ulrich+Horne:1996,Kollatschny:2003,Bentz+:2010,Grier+:2013}. Though, a leading blue side has been observed in some sources \citep[e.g.,][]{Denney+:2009}. In the simplest outflowing disk-wind model, it is expected that there will be little lag in the blue line wing compared to the variation in the continuum, while the red wing experiences up to twice the delay \citep{Gaskell:2009}. \citet{Gaskell+Goosmann:2016} suggested that the shorter response in the red wing is an evidence of inflow motion, and hence disfavouring the outflowing wind model.
  
  However, \citet{Mangham+:2017} using detailed radiative transfer and ionisation treatment of a disk-wind simulation argued that the classical indicator of Keplerian rotation, inflow, or outflow is not always denoted by the symmetric, red wing leading blue wing, or blue wing leading red wing line profile signature. Even for flow dominated by a rotational velocity component \citep{Chiang+Murray:1996,Kashi+:2013,Waters+:2016,Mangham+:2017}, a faster response in the red wing, which is often considered as a signature of inflow, can be produced in moderate luminosity objects \citep{Mangham+:2017}. An outflow signature is only apparent in high luminosity objects \citep{Mangham+:2017}.
  
  \citet{Chiang+Murray:1996} demonstrated that an earlier time delay of the red wing can be achieved in their outflowing spherical wind model due to the contributions of radial and rotational velocity components from the radiative transfer effects. By modelling a narrow outflowing wind model, \citet{Yong+:2017} have also found quicker response in the red side of the line profile in some regions of the wind especially for face-on inclination angle. Furthermore, the dusty failed wind of \citet{Czerny+Hryniewicz:2011} and quasar rain model of \citet{Elvis:2017} suggest that both infall and outflow are present in the wind. This is consistent with the observations from dynamical modelling of the BLR \citep{Pancoast+:2014} and velocity-resolved reverberation mapping \citep{Grier+:2013}.

  \item \textbf{Radio loudness and BAL.}\\
  The presence of both BALs and radio jets in a quasar was initially believed to be impossible \citep[e.g.,][]{Stocke+:1992}. However, numerous radio-loud BALQs now been detected \citep[e.g.,][]{Becker+:1997,Brotherton+:1998}. Yet, only $\sim 5$ per cent of radio-loud quasars are BALQs \citep[e.g.,][]{Becker+:2001,Menou+:2001}, compared to the $\sim 15$ per cent of the entire quasar population being BALQs \citep[e.g.,][]{Kellermann+:1989,Ivezic+:2002,Balokovic+:2012}. BALQs with high radio power and radio luminosity are even rarer \citep{Gregg+:2006,Shankar+:2008}. This discrepancy can be explained in the context of the orientation scenario, due to relativistic beaming of the radio jets along the observer's line-of-sight. In this case, a fraction of radio-quiet quasars will be relativistically beamed towards the observer and boosted to higher radio luminosities. These beamed radio sources become a disproportionate fraction of the bright radio source. This reasoning is discussed in \citet{Shankar+:2008}, though they also pointed out that other mechanisms are necessary to accelerate the BAL winds in polar radio-loud quasars.
  
  In our model, we have an ionisation cone that is aligned with the radio jet. An observer looking down this ionisation cone will not observe BALs. Therefore, BALQs will generally be non-beamed sources, and even when BALQs are a fixed fraction of radio quasars, the observed fraction of BALQs seen will decrease with radio power due to the apparent increase in radio-loud quasars numbers due to beaming.
  
  Using samples in the optical domain, \citet{Bruni+:2014} reported that the BALQs have similar geometries, black hole masses, and accretion rates, regardless of their radio properties. \citet{DiPompeo+:2012b,Rochais+:2014} also found no statistical difference in the BAL features between radio-loud and radio-quiet ultraviolet BALQ spectra. This hints that the radio loudness might be independent of the BAL characteristics. Thus, the findings can be generalised to both radio phases of BALQs \citet{Rochais+:2014}.
  
  Our model allows a wide range of BAL viewing angles, and therefore a range of observed radio properties. We also do not expect the radio loudness of the AGN to affect the BLR properties, hence we expect the observed similarities between the radio-loud and radio-quiet samples.

  \item \textbf{X-ray observational features.}\\
  X-ray spectra of AGNs have revealed the presence of ionised absorption, which is often considered to be indicative of outflowing photoionised material along the line-of-sight \citep{Halpern:1984}. The X-ray blueshifted absorption lines, referred to as warm absorbers (WAs), are associated to highly ionised gas with velocity of order 100--1000\,\kms\ \citep[e.g.,][]{Kaastra+:2000,Kaspi+:2002,McKernan+:2007}. WAs are also common and present in about half of Seyfert~1 galaxies \citep[e.g.,][]{Reynolds+Fabian:1995,Reynolds:1997,George+:1998}. Although less is known about the exact nature and origin of the WAs \citep[see review by e.g.,][]{Crenshaw+:2003}, several suggestions include evaporating clouds in the BLR \citep[e.g.,][]{Netzer:1996}, scattering gas through the obscuring torus \citep[e.g.,][]{Krolik+Kriss:1995}, two-component WA regions \citep[e.g.,][]{Otani+:1996}, and accretion disk wind \citep[e.g.,][]{Konigl+Kartje:1994,Elvis:2000,Bottorff+:2000}. There is also an extreme class of absorbers, called ultra-fast outflows (UFOs), with velocity  $\gtrsim 10^{3}\,\kms$ and even extend up to $\sim 0.4\,c$, where $c$ is the speed of light \citep[e.g.,][]{Chartas+:2002,Reeves+:2003,Chartas+:2003,Pounds+:2003}. It has been suggested that the UFOs and WAs belong to the same single large-scale stratified wind, with WAs located further away from the black hole than the UFOs \citep[e.g.,][but see also \citealt{Laha+:2014}]{Tombesi+:2013}. We do not attempt to explain these phenomena in our model.

  \item \textbf{Unification of AGNs.}\\
  The basis of our proposed BLR disk-wind model is based on the study of BAL features from quasar samples. However, to a certain degree, we believe that our model could be extended to other types of AGNs, such as Seyfert galaxies. Although, the strength of the outflow is likely to be dependent on the properties of the AGN, in a sense that more luminous systems are associated with stronger BAL winds while narrower absorptions occur in less luminous systems \citep{Laor+Brandt:2002,Ganguly+:2007}. Since Seyferts are low or moderate luminosity counterparts of quasars, those with BALs are scarce and only a transient BAL outflow, for example in the narrow-line Seyfert~1 galaxy WPVS~007 \citep{Leighly+:2009}, is found. In this case, we expect Seyferts to fit into our model but with wind power scaled to its luminosity and black hole mass.
\end{itemize}

Our model differs from the majority of disk-wind models as it does not have a narrow opening angle for the wind. Additionally, the wind is stratified in both ionisation and probably density by angle from the axis of accretion and distance from the central accretion disk. The model quite naturally accounts for the substantial observed similarities between BALQs and non-BALQs, while retaining some of the key elements of the models of \citet{Murray+:1995,Elvis:2000,Elvis:2004}.

\section{Summary} \label{sec:summary}

Using quasar samples from the SDSS DR12Q, we have investigated the various characteristics of the continuum and BELs, namely the absolute magnitude, redshift spectral index, FWHM, asymmetry, EW, and velocity shifts for HIL \ion{C}{iv} and LIL \ion{Mg}{ii}. We have applied statistical tests and supervised machine learning for classification to examine whether the attributes for BALQ and non-BALQ populations originate from the same parent population.

Although a few parameters have shown statistical differences between BALQ and non-BALQ samples, the overriding result is that the two populations have largely similar properties. The shape of the distributions are highly similar in most cases. Analysis from machine learning also points out the complexity in separating the two classes as all the algorithms employed only performed marginally in the classification tasks. These observed trends appear to be inconsistent with predictions from a purely orientation explanation based on a narrow disk-wind for explaining the BALQ population. We have no evidence against an evolutionary model.

Under the assumption that the BAL and BEL regions are co-spatial, our analysis can be use to infer something about the structure of the BLR. BELs are standard features of quasars from almost all lines-of-sight that are not obscured by a dusty torus. In the orientation explanation, the BLR is described as a traditional disk-wind model that is constrained within a narrow angular range. This means that the covering factor of the absorbing part of the wind is reflected by the proportion of BALQs. Importantly, the measured profiles of BELs will vary significantly with line-of-sight. Since we do not observe any significant difference between the properties of the BALQs and non-BALQs, BALs must not have a preferred direction. This argues against a disk-wind of small opening angle, and instead favours a clumpy wind covering a wide range of angles.

A modified model for the disk-wind is required and proposed, retaining key features of the traditional models:
\begin{itemize}[leftmargin=.125in,listparindent=\parindent,itemsep=1ex]
  \item The clumpy wind is stratified and covers a wide range of angles.
  \item The wind consists of multiple radial streams of high density gas and BALs are seen when the viewing angle intersects with these streams in the wind.
  \item The high-ionisation lines, such as \ion{C}{iv}, lie close to the ionising source, while low-ionisation lines, such as \ion{Mg}{ii}, lie further from the source but close to the wind streamline.
\end{itemize}
This model appears to consistently explain a lot of the observed features in BAL and non-BAL quasars.

\section*{Acknowledgements}

We thank the anonymous referee for valuable suggestions on the manuscript. NFB thanks the STFC for support under Ernest Rutherford Grant ST/M003914/1. This research has made use of the VizieR catalog access tool, CDS, Strasbourg, France. The original description of the VizieR service was published in \citep{Ochsenbein+:2000}. This research made use of the Python libraries including open source packages such as \texttt{astropy} \citep{Astropy:2013}, \texttt{ipython} \citep{Perez+:2007}, \texttt{matplotlib} \citep{Hunter:2007}, \texttt{numpy} \citep{vanderWalt+:2011}, \texttt{pandas} \citep{McKinney:2010}, \texttt{scikit-learn} \citep{Pedregosa+:2011}, and \texttt{scipy} \citep{Jones+:2001}.


\bsp	
\label{lastpage}
\end{document}